 \documentclass[final,3p,times,twocolumn]{elsarticle}

\usepackage{amsmath,amssymb}
\usepackage{tikz}
\usepackage{braket}
\usepackage[colorlinks, linkcolor=red, anchorcolor=green, citecolor=blue]{hyperref}

\usepackage{mcode}
\usepackage{algorithm,algorithmic}
\usepackage{listings}
\usepackage[percent]{overpic}
\usepackage{float}
\usepackage{comment}
\usepackage{verbatim}
\usepackage{indentfirst}

\newcommand{\vk}{\mathbf{k}}
\newcommand{\vg}{\mathbf{g}}
\newcommand{\br}{\mathbf{r}}
\newcommand{\brp}{\mathbf{r^{\prime}}}

\newcommand{\Etot}{E_{\mathrm{tot}}}
\newcommand{\Eion}{E_{\mathrm{ion}}}
\newcommand{\Ekin}{E_{\mathrm{kin}}}
\newcommand{\Ehartree}{E_{\mathrm{Hartree}}}
\newcommand{\Exc}{E_{\mathrm{xc}}}
\newcommand{\Enuc}{E_{\mathrm{nuc}}}
\newcommand{\Vion}{\hat{V}_{\mathrm{ion}}}
\newcommand{\Vhartree}{\hat{V}_{\mathrm{Hartree}}}
\newcommand{\Vxc}{\hat{V}_{\mathrm{xc}}}
\newcommand{\Vhse}{\hat{V}_{\mathrm{x}}^{\mathrm{HSE}}}
\newcommand{\Vace}{\hat{V}_{\mathrm{x}}^{\mathrm{ACE}}}
\newcommand{\Phiref}{\Phi_{\mathrm{ref}}}

\newcommand{\eg}{{\it e.g.}}
\newcommand{\ie}{{\it i.e.}}

\newcommand{\kssolv}{KSSOLV}
\newcommand{\matlab}{MATLAB}

\newcommand{\CY}[1]{{\color{red}~\textsf{[CY: #1]}}}

\journal{Computer Physics Communication}

\begin{document}

\begin{frontmatter}



\title{KSSOLV 2.0: An efficient MATLAB toolbox for solving the Kohn-Sham equations with plane-wave basis set\tnoteref{t1}}

\tnotetext[t1]{This paper and its associated computer program are available via the Computer Physics Communications homepage on ScienceDirect }

\author[1]{Shizhe Jiao}
\author[1]{Zhenlin Zhang}
\author[1]{Kai Wu}
\author[1]{Lingyun Wan}
\author[1]{Huanhuan Ma}
\author[1]{Jielan Li}
\author[1]{Sheng Chen}
\author[1]{Xinming Qin}
\author[1]{Jie Liu}
\author[1]{Zijing Ding}
\author[1]{Jinlong Yang}
\address[1]{School of Future Technology, Department of Chemical Physics, and Anhui Center for Applied Mathematics, University of Science and Technology of China, Jinzhai Road No.96, Hefei, 230026, PR China}

\author[2]{Yingzhou Li\corref{cor1}}
\ead{yingzhouli@fudan.edu.cn}
\address[2]{School of Mathematical Sciences, Fudan University, Handan Road No.220, Shanghai, 200433, PR China}

\author[1]{Wei Hu\corref{cor1}}
\ead{whuustc@ustc.edu.cn}

\author[3,4]{Lin Lin\corref{cor1}}
\ead{linlin@math.berkeley.edu}
\address[3]{Department of Mathematics, University of California, Berkeley, CA 94720, USA}
\address[4]{Applied Mathematics and Computational Research Division, Lawrence Berkeley National Laboratory, Berkeley, CA 94720, USA}

\author[4]{Chao Yang\corref{cor1}}
\ead{chaoyang@lbl.gov}
\cortext[cor1]{Corresponding author.}

\begin{abstract}

KSSOLV (Kohn-Sham Solver) is a MATLAB  toolbox for performing Kohn-Sham density functional theory (DFT) calculations with a plane-wave basis set. KSSOLV 2.0 preserves the design features of the original KSSOLV software to allow users and developers to easily set up a problem and perform ground-state calculations as well as to prototype and test new algorithms. Furthermore, it includes new functionalities such as new iterative diagonalization algorithms, k-point sampling for electron band structures, geometry optimization and advanced algorithms for performing DFT calculations with local, semi-local, and hybrid exchange-correlation functionals. It can be used to study the electronic structures of both molecules and solids. We describe these new capabilities in this work through a few use cases. We also demonstrate the numerical accuracy and computational efficiency of KSSOLV on a variety of examples.

\end{abstract}




\end{frontmatter}

\textbf{Program Summary}

\textit{Program title:} Kohn-Sham Solver 2.0 (KSSOLV 2.0)

\textit{CPC Library link to program files:}

\textit{Developer's repository link:} https://bitbucket.org/berkeleylab/kssolv2.0/src/release/

\textit{Licensing provisions:} BSD

\textit{Programming language:} MATLAB

\textit{Nature of problem:} KSSOLV2.0 is used to perform Kohn-Sham density functional theory based electronic structure calculations to study chemical and material properties of molecules and solids. The key problem to be solved is a constrained energy minimization problem, which can also be formulated as a nonlinear eigenvalue problem.

\textit{Solution method:} The KSSOLV 2.0 implements both the self-consistent field (SCF) iteration with a variety of acceleration strategies and a direct constrained minimization algorithms.  It is written completely in MATLAB and uses MATLAB's object oriented programming features to make it easy to use and modify.


\section{Introduction} \label{sec:Introduction}

KSSOLV (Kohn-Sham Solver)~\cite{yang2009kssolv} is a MATLAB (Matrix Laboratory) toolbox for performing Kohn-sham density functional theory (DFT)~\cite{HohenbergKohn1964, kohn1965} based electronic structure calculations. It uses the plane-wave basis set to represent electron wavefunctions. One of the original motivations for developing such a software package was to make it easy to prototype and test new algorithms for solving the Kohn-Sham nonlinear eigenvalue problems. KSSOLV leverages the high quality numerical linear algebra functions and object-oriented features of MATLAB to enable researchers who have minimal knowledge of other electronic structure calculation software written in FORTRAN or C/C++ to quickly modify existing algorithms, as well as to develop and test new ideas. Over the last decade, KSSOLV has become a useful research and teaching tool for studying electronic structures of molecules and solids and developing new methods for solving the Kohn-Sham problem, as evidenced by an increasing number of publications that use KSSOLV to perform numerical experiments required to demonstrate improved convergence or better accuracy of new theoretical methods and numerical algorithms. Examples of such developments include tensor hypercontraction algorithm~\cite{kssolvcite1-1}, improved generalized Davidson algorithm~\cite{kssolvcite1-2}, elliptic preconditioner in self-consistent field iteration~\cite{ksslovcite1-3}, low rank approximation in G$_{0}$W$_{0}$ calculations~\cite{kssolvcite1-4}, perturbation theory~\cite{kssolvcite2-3}, quantum embedding theory~\cite{kssolvcite2-4}, quantum computation~\cite{kssolvcite2-6},  linear-response time-dependent density functional theory~\cite{kssolvcite2-2, kssolvcite2-8},  phonon calculations~\cite{LinXuYing2017}, proximal gradient method~\cite{zaiwen1}, trace-penalty minimization method~\cite{zaiwen2}.



As KSSOLV becomes more widely used, it also becomes clear that the functionalities supported in the original KSSOLV software package are insufficient.  For example, the original KSSOLV could only be used to perform single-point calculations of the ground state energies for molecules placed in a large supercell, and only the local density approximation (LDA) exchange-correlation functional was implemented. Moreover, the pseudopotentials supported in the original KSSOLV did not clearly specify the type of pseudopotentials used and did not allow widely accepted pseudopotential libraries to be easily incorporated.
The lack of these functionalities makes the comparison of KSSOLV with other software packages somewhat difficult. 

To address these issues, we have recently revamped the development of KSSOLV by adopting more standard pseudopotential types, such as the ONCV (Optimized Norm-Conserving Vanderbilt)~\cite{oncv} and Hartwigsen-Goedecker-Hutter(HGH)~\cite{hgh} pseudopotentials. We have incorporated more recent algorithmic development, and added new functionalities and features without sacrificing the usability of the software. The new software release, KSSOLV 2.0, is an open source software\footnote{Bitbucket repository with documentation: \url{https://bitbucket.org/berkeleylab/kssolv2.0/src/release/}.}. In addition to being a flexible tool for new algorithm development, KSSOLV 2.0 can also be easily used to study the properties of molecules and solids. It serves as both a research and teaching tool for researchers engaged in the simulation and prediction of chemical and material properties, such as linear-response time-dependent density functional theory~\cite{kssolvcite2-2}, many-electron self energy calculations~\cite{kssolvcite2-1}, structure optimization~\cite{kssolvcite2-5}, photocatalytic materials simulations~\cite{kssolvcite2-7}.

We strive to make KSSOLV 2.0 as efficient as possible without sacrificing its readability and usability. In addition to traditional platforms, the software can also be run on heterogeneous architectures with graphics processing units (GPUs)~\cite{kssolvgpu}.
However, KSSOLV 2.0 is not designed for performing large-scale electronic structure calculations. For these types of calculations, many existing software tools can be used alternatively, such as 
Gaussian~\cite{g16}, NWChem (NorthWest computational Chemistry)~\cite{valiev2010nwchem}, Q-CHEM~\cite{shao2015advances}, BDF (Beijing Density Functional program package)~\cite{zhang2020bdf}, and PySCF (Python-based Simulations of Chemistry Framework)~\cite{sun2018pyscf} within Gaussian-type orbital (GTO) basis set; SIESTA (Spanish Initiative for Electronic Simulations with Thousands of Atoms)~\cite{soler2002siesta},
HONPAS (Hefei Order-N Packages for Ab initio Simulations)~\cite{qin2015honpas,xiang2006linear,qin2020interpolative}, FHI-aims (Fritz Haber Institute ab initio molecular simulations)~\cite{blum2009ab} and ABACUS (Atomic-orbital Based Ab-initio Computation at Ustc)~\cite{chen2010systematically} within numerical atomic orbital (NAO) basis set; and VASP (Vienna Ab initio Simulation Package)~\cite{kresse1993ab}, ABINIT~\cite{gonze2002first}, QE (QUANTUM ESPRESSO)~\cite{giannozzi2009quantum},  PWmat~\cite{jia2013fast}, PWDFT (Plane-Wave Density Functional Theory)~\cite{hu2017adaptively} within plane-wave basis set. These DFT codes are often written in languages such as FORTRAN and C++, and parallelized with OpenMP, MPI, and CUDA. The compilation, installation and usage of these software packages often take a significant amount of effort. Our software is more similar to some other recently developed DFT toolboxes such as GPAW(Grid-based Projector Augmented Wave)~\cite{mortensen2005real,enkovaara2010electronic}, M-SPARC (Matlab-Simulation Package for Ab-initio Real-space Calculations)~\cite{xu2020m} and PWDFT.jl~\cite{fathurrahman2020pwdft}, DFT.jl~\cite{dftk}, which are based on higher-level scripting languages such as Python, Julia, and MATLAB. We should point out that the M-SPARC software, which is written in MATLAB, focuses on a real space discretization of the Kohn-Sham problem whereas KSSOLV uses a plane-wave discretization. The main characteristics of these software packages are shown in Table~\ref{allsoft}. The advantage of KSSOLV is that it is written completely in MATLAB, which is designed to perform linear algebra operations in a straightforward manner.
MATLAB also provides an excellent Integrated Development Environment(IDE), which makes the development process much easier than other software tools. Furthermore, the unique profiling capability of MATLAB allows us to easily identify computational bottlenecks.

\begin{table*}[!tb]\footnotesize
\centering
\setlength{\tabcolsep}{3mm}
\caption{The characteristics of several DFT software packages, including programming language, basis, license, language type and publish year. GTOs: gaussian-type orbital (GTO) basis set, NAOs: numerical atomic orbital basis set, PW: plane-wave basis set. AE: all electronic calculation, NCPPs: norm-conserving pseudopotentials, ECP: effective core potential, PAW: projector augmented wave. 
GPL: GNU General Public License, ECL-2.0: Educational Community License, BSD: Berkeley Software Distribution License.} 
\label{Parameters}
\begin{tabular}{cccccccc} \ \\
\hline \hline
Software   & Language  & Basis & AE/PSP          & License    & Language type & Year  &  Reference  \ \\
\hline
Gaussian   &  Fortran  &  GTOs & AE/ECP              & Commercial  & Compiled language       & 1970  &  \cite{g16} \ \\
NWChem     &  Fortran  &  GTOs/PW & AE/PAW              & Free, ECL-2.0        & Compiled language        & 1994  & \cite{valiev2010nwchem} \ \\
QChem      &  Fortran  &  GTOs & AE/ECP              & Academic, commercial        & Compiled language       & 1997  & \cite{shao2015advances} \ \\
BDF        &  Fortran  &  GTOs & AE              & Free, GPL        & Compiled language       & 2009  & \cite{zhang2020bdf} \ \\
\hline
SIESTA     &  Fortran  &  NAOs & NCPPs             & Free, GPL        & Compiled language        & 1996  & \cite{soler2002siesta} \ \\
HONPAS     &  Fortran  &  NAOs & NCPPs             & Free, GPL        & Compiled language        & 2005  & \cite{qin2015honpas} \ \\
FHI-aims   &  Fortran  &  NAOs & AE              & Academic, commercial  & Compiled language        & 2009  & \cite{blum2009ab} \ \\
ABACUS     &  Fortran  &  NAOs/PW & NCPPs       & Free, GPL        & Compiled language        & 2016  & \cite{chen2010systematically} \ \\
\hline
VASP       &  Fortran  &  PW   & PAW             & Commercial  & Compiled language        & 1989  & \cite{kresse1993ab} \ \\
ABINIT     &  Fortran  &  PW   & NCPPs/PAW       & Free, GPL        & Compiled language        & 1998  & \cite{gonze2002first} \ \\
QE         &  Fortran  &  PW   & NCPPs/PAW      & Free, GPL        & Compiled language        & 2001  & \cite{giannozzi2009quantum} \ \\
PWmat      &  Fortran  &  PW   & NCPPs       & Commercial  & Compiled language        & 2013  & \cite{jia2013fast} \ \\
PWDFT      &  C/C++    &  PW   & NCPPs       & Free, BSD        & Compiled language        & 2017  &  \cite{hu2017adaptively} \ \\
\hline
GPAW       &  Python   &  PW   & PAW             & Free, GPL        & Interpreted language     & 2003  & \cite{mortensen2005real} \ \\
KSSOLV     &  MATLAB   &  PW   & NCPPs       & Free, BSD        & Interpreted language   & 2009  & \cite{yang2009kssolv}; This work \ \\
PySCF      &  Python   &  GTOs  & AE; NCPPs        & Free, BSD        & Interpreted language      & 2014  & \cite{sun2018pyscf} \ \\
M-SPARC    &  MATLAB   &  RS   & NCPPs             & Free, GPL        & Interpreted language         & 2019  &  \cite{xu2020m} \ \\
PWDFT.jl   &  Julia    &  PW   & NCPPs             & Free, GPL        & Interpreted language        & 2020  &  \cite{fathurrahman2020pwdft} \ \\
DFT.jl   &  Julia    &  PW   & NCPPs             & Free, GPL        & Interpreted language        & 2021  &  \cite{dftk} \ \\
\hline \hline
\end{tabular}
\label{allsoft}
\end{table*}

This work is organized as follows. In the next section, we briefly summarize the main methodology and standard methods implemented in KSSOLV for solving the Kohn-Sham DFT problem, as well as a number of recently developed and more advanced algorithms.  We highlight the object-oriented design feature of KSSOLV in section~\ref{sec:design}, and demonstrate several main features of KSSOLV 2.0 through a number of use cases in section~\ref{sec:usecase}.  The accuracy and efficiency of KSSOLV 2.0 are reported in section~\ref{sec:results} for several small to medium sized benchmark test problems.

\section{Methodology} \label{sec:Methodology}

KSSOLV 2.0 is designed to perform Kohn-Sham density functional theory (KS-DFT) based electronic structure calculations. In this section, we
briefly describe the main mathematical problem to be solved, namely, the
Kohn-Sham nonlinear eigenvalue problem, or equivalently, the Kohn-Sham
total energy minimization problem. In KSSOLV 2.0, the eigenfunction to be
computed is expanded in a plane-wave basis, which will be discussed briefly
in section~\ref{sec:planewave}. A key component of the Kohn-Sham
Hamiltonian operator is the exchange-correlation (XC) potential that
accounts for many-body effects in a many-electron system. We describe the XC functions implemented in KSSOLV 2.0 in section~\ref{sec:exc}. KSSOLV 2.0 employs the pseudopotential
method which is commonly used to address the weak singularity (cusp) in
the nuclei-electron potential. We briefly discuss pseudopotentials used in
KSSOLV 2.0 in section~\ref{sec:pseudo}. 
Sections~\ref{sec:scfdcm} and ~\ref{sec:eigen} 
are concerned with several numerical algorithms used in KSSOLV 2.0 to solve
the Kohn-Sham and related problems. In particular, we discuss new
algorithms that have been added in the latest release of KSSOLV 2.0 in sections~\ref{sec:relaxatom},~\ref{sec:scdm} and~\ref{sec:C}.

\subsection{Mathematical formulation}\label{sec:basic}

\subsubsection{Brief introduction of KS-DFT}\label{sec:ksdft}

The KS-DFT~\cite{HohenbergKohn1964, kohn1965} is the most widely used
methodology to perform first-principles calculations and materials
simulations to study the electronic structure of molecules and solids.

The key problem to be solved in KS-DFT based electronic structure
calculation of an atomistic system with $N_e$ electrons is a nonlinear
eigenvalue problem of the form
\begin{equation}
\hat{H}(\rho) \psi_j =  \varepsilon_j \psi_j,
\label{eq:kseig}
\end{equation}
where $j = 1,2,...,N_e$, $\varepsilon_1 \leq \varepsilon_2 \leq \cdots
\leq \varepsilon_{N_e}$ are $N_e$  eigenvalues of $\hat{H}(\rho)$. They
are known as the Kohn-Sham eigenvalues associated with the corresponding eigenfunctions $\psi_j$'s, also known as the \textit{occupied} orbitals or states. The function $\rho$ is the electron density defined (at zero temperature) as
\begin{equation}
\rho = \sum_{j=1}^{N_e} |\psi_j|^2.
\label{eq:rho}
\end{equation}

The Kohn-Sham Hamiltonian $\hat{H}$ to be partially diagonalized is a functional of  $\rho$ (and consequently $\psi_j$'s.)

Equation \eqref{eq:kseig} is the first order necessary condition
associated with a constrained minimization problem
\begin{equation}
\min_{\braket{\psi_i,\psi_j}=\delta_{i,j}} E_{\rm{tot}}(\{\psi_i\}),
\label{eq:etotmin}
\end{equation}
where $E_{\rm{tot}}$ is a total energy functional that consists of both
kinetic and various potential terms~\cite{kohn1965}, i.e.
\begin{equation}
E_{\rm{tot}}=\Ekin+\Ehartree+\Eion+\Enuc+\Exc,
\label{eq:energy}
\end{equation}
where $\Ekin$ represents the kinetic energy, $\Ehartree$ is the potential
energy induced by electron-electron repulsion, $\Eion$ is the potential
energy induced by nucleus-electron attraction, $\Enuc$ is the potential
energy induced by nucleus-nucleus repulsion, and $\Exc$ is the
exchange-correlation energy that accounts for the many-body effects unaccounted in the preceding terms.  The mathematical expressions for
these energy terms can be found in the standard literature~\cite{energyform,martin2020electronic}.

As a result, the Kohn-Sham Hamiltonian $\hat{H}$ appeared in
\eqref{eq:kseig} can be partitioned accordingly, i.e.
\begin{equation}
\hat{H}= \hat{T}+\Vhartree+\Vion+\Vxc,
\label{eq:ksham}
\end{equation}
where $\hat{T}$ is the kinetic energy operator, $\Vion$ is the ionic
potential operator, $\Vhartree$ is the Hartree potential operator and
$\Vxc$ is the exchange-correlation potential operator. We again refer readers to standard literature~\cite{martin2020electronic} for analytical expressions for each one of these terms.

\subsubsection{Plane-wave basis set}\label{sec:planewave}

The Kohn-Sham Hamiltonian is periodic for solids. For such  periodic systems, we solve \eqref{eq:kseig} by focusing on one period, often known as a primitive cell. 
It follows from the Bloch’s theory that a Kohn-Sham orbital $\psi_j(\br)$ takes the form
\begin{equation}
\psi_{j,\vk}=e^{i \textbf{k} \br} u_{j,\vk}(\br),
\end{equation}
where $u_{j,\vk}(\br)$ is periodic and $\vk$ is a crystal momentum vector in the first Brillouin zone.

The occupied Kohn-Sham orbitals are indexed by both $j$ and $\vk$.
The charge density is periodic and defined as
\begin{equation}
\rho(\br)=\frac{|\Omega|}{(2 \pi)^{3}} \int_{B Z} \rho_{\vk}(\br) d \vk,
\label{eq:rhok}
\end{equation}
where $|\Omega|$ is the volume of the primitive cell in the real space, and 
\[
\rho_{\vk}(\br)=\sum_{j=1}^{N_{e}}\left|\psi_{j, \vk}(\br)\right|^{2},
\]

The choice of a periodic unit cell is not unique. When the unit cell is
sufficiently large in real space, the corresponding unit cell in first
Brillouin zone is so small that the integral in \eqref{eq:rhok} can be
approximated the evaluation of $\rho_{\vk}(\br)$ at a single $\vk$-point,
e.g., $\vk=0$, also known as the $\Gamma$-point.

Because $u_{j,\vk}$ is periodic, it can be expanded by plane-wave basis functions,
i.e.,
\begin{equation}
u_{j,\vk}(\br)=\sum_{\ell=1}^{\infty} c_{j,\ell}^{\vk} e^{i \vg_{\ell}^T \br}, \ \
c_{j,\ell}^{\vk}=\int_{\Omega} u_{j,\vk}(\br) e^{-i \vg_{\ell}^T \br} d \br,
\label{eq:fourier}
\end{equation}
where $\vg_{\ell}$ is a lattice vector in the reciprocal space.  As a
result, a Kohn-Sham orbital $\psi_{j,\vk}$ can be represented
by
\[
\psi_{j,\vk} = \sum_{\ell=1}^{\infty} c_{j,\ell}^{\vk} e^{i (\vk + \vg_{\ell})^T \br}. \ \
\]

This is the discretization scheme used in KSSOLV as in other plane-wave based electronic structure calculation software packages.

In practice, the infinite sum in \eqref{eq:fourier} is truncated and
approximated by a finite sum. As in all other plane-wave based Kohn-Sham
solvers, the truncation of the plane-wave expansion is based on the
following criterion
\begin{equation}
\left| \vk + \vg_{\ell}\right|^{2}<2 E_\textrm{cut},
\end{equation}
for some energy cut-off value $E_{\rm{cut}}$. If the number of $\vg$'s that
satisfy this criterion is $N_g$,  an approximation to the Kohn-Sham
orbital $\psi_{j,\vk}$ can be written as
\begin{equation}
\psi_{j,\vk}(\br)\approx\sum_{\ell=1}^{N_{g}} c_{j,\ell}^{\vk} e^{i (\vk + \vg_{j})^T \br}.
\end{equation}

In a plane-wave basis set, the representations of $\hat{T}$ and $\Vhartree$ in
\eqref{eq:ksham} are particularly simple, i.e., they are diagonal (or
local). However, $\Vion$ and $\Vxc$ typically have a more compact
representation in real space. As a result, when $\psi_{j,\vk}$ are
discretized by a plane-wave expansion, the Kohn-Sham Hamiltonian $\hat{H}$
is not constructed or stored explicitly. The multiplication of $\hat{H}$ (which is called an implicit Hamiltonian)
with $\psi_{j,\vk}$ can be implemented efficiently by working with both the
real space and reciprocal space representations of $\psi_{j,\vk}$. The
change of representation between real space and reciprocal space is
facilitated by Fast Fourier Transforms (FFTs). This is a key feature of
plane-wave based Kohn-Sham equation solver.

\subsubsection{Exchange-correlation functional}\label{sec:exc}

The exchange-correlation energy term $\Exc$ in~\eqref{eq:energy} and the
exchange-correlation Hamiltonian term $\Vxc$ accounts for the many-body
effects of electron interactions. They are particularly significant for
KS-DFT. The exact analytical forms of $\Exc$ and $\Vxc$ are unknown.
Various approximations have been proposed. These include the local density
approximation (LDA)~\cite{perdew1981self}, generalized gradient
approximation (GGA)~\cite{perdew1996generalized}, and the hybrid
functional~\cite{stroppa2008shortcomings, schimka2010accurate,
sun2015comparing}. A hybrid functional includes a fraction of the exact exchange
potential from the Hartree-Fock (HF)~\cite{baerends1973self} theory.  Three widely used hybrid functionals are shown in~\eqref{eq:hybridxc}. In
KSSOLV 2.0, all three approximations have been implemented. Both LDA and GGA are local in real space. Hence, applying these potential operators to a wavefunction is relatively straightforward.  However, the Hartree-Fock exact
exchange term in a hybrid functional is nonlocal, and applying it to a wavefunction is more costly.  However, efficient methods for applying this
term have been developed~\cite{JCTC_12_2242_2016_ACE, hu2017projected,
hu2017interpolative, lee2019systematically}. We will discuss efficient
methods for working with hybrid functional KS-DFT in section~\ref{sec:C}.
\begin{equation}
\begin{split}
E_{\mathrm{xc}}^{\mathrm{PBE}0} = {}&\frac{1}{4} E_{\mathrm{x}}^{\mathrm{HF}}+\frac{3}{4} E_{\mathrm{x}}^{\mathrm{PBE}}+E_{\mathrm{c}}^{\mathrm{PBE}} \ \\
E_{\mathrm{xc}}^{\mathrm{HSE}} = {}& 0.25 E_{\mathrm{x}}^{\mathrm{HF}, \mathrm{SR}} + 0.75 E_{\mathrm{x}}^{\mathrm{PBE}, \mathrm{SR}}+E_{\mathrm{x}}^{\mathrm{PBE}, \mathrm{LR}}+E_{\mathrm{c}}^{\mathrm{PBE}}\ \\
E_{\mathrm{xc}}^{\mathrm{B} 3 \mathrm{LYP}} = {}& E_{\mathrm{x}}^{\mathrm{LDA}}+0.2\left(E_{\mathrm{x}}^{\mathrm{HF}}-E_{\mathrm{x}}^{\mathrm{LDA}}\right)+0.72\left(E_{\mathrm{x}}^{\mathrm{GGA}}-\right.\\
& E_{\mathrm{x}}^{\mathrm{LDA}})+E_{\mathrm{c}}^{\mathrm{LDA}}+0.81\left(E_{\mathrm{c}}^{\mathrm{GGA}}-E_{\mathrm{c}}^{\mathrm{LDA}}\right),
\end{split}
\label{eq:hybridxc}
\end{equation}

\subsubsection{Pseudopotential}\label{sec:pseudo}

\kssolv{} adopts the pseudopotential methodology~\cite{pseupotential} to model the interaction between nuclei and electrons.
In this approach, core electrons are treated as a part of an ionic core represented by a pre-computed effective potential.  Only the valence electrons are present in \eqref{eq:kseig} and \eqref{eq:rho}. For a plane-wave DFT code, the
pseudopotential method allows us to significantly reduce the computational cost by reducing the number of active electrons
and the number of planewaves required to represent eigenfunctions of the Kohn-Sham Hamiltonian. The latter reduction is due to the fact that the use of pseudopotential makes the eigenfunction of the corresponding Kohn-Sham Hamiltonian less oscillatory.

There are two common types of pseudopotentials in modern DFT computation,
namely norm-conserving pseudopotential~(NCPP) and ultrasoft
pseudopotential. 
In general,  the implementation of NCPPs is easier than that for ultrasoft pseudopotentials~\cite{PPs1,PPs2},
and they produce sufficient accuracy for many systems.
Therefore, NCPPs are the supported type of pseudopotentials in \kssolv{}.

A pseudopotential typically consists of a local component $V_{\mathrm{loc}}(\br)$ and a nonlocal component $V_{\mathrm{NL}}(\br,\br')$. By using the Kleinman-Bylander form of an atomic pseudopotential, we can express $V_{\mathrm{NL}}(\br,\br')$ in a low rank separable form
\begin{equation} \label{eq:kbpp}
V_{\mathrm{NL}}(\br,\br') =
\sum_{lm} \beta_{lm}(\br) v_l \beta_{lm}(\br')^{*},
\end{equation}
where $\beta_{lm}(\br)$ is a pseudo atomic wavefunction associated with the quantum numbers $l$ and $m$, and $v_l$ is a weighting factor that depends on the degree of spherical harmonic used in $\beta_{lm}$. 

There are many ways to construct pseudopotentials. We refer readers to standard literature on this subject~\cite{martin2020electronic},  like many other KS-DFT software tools, we use pseudopotentials archived at a URL\footnote{pseudopotentials homepage used by \kssolv{}: \url{http://pseudopotentials.quantum-espresso.org/legacy_tables}.} in KSSOLV. 
KSSOLV 2.0 can read pseudopotential files in multiple formats, and convert them to suitable real or reciprocal space representations.
The local and non-local components are treated differently. The local component is represented in real space and applied as a diagonal matrix. It is constructed by the summing local atomic potentials re-centered at atomic positions. The re-centering and the summation are carried out through Fourier transforms. The nonlocal pseudo wavefunctions are stored and applied in the
reciprocal space. For atoms of the same type, their nonlocal pseudo wavefunctions are combined and transformed to reciprocal space via
spherical harmonic transform. The pseudo wavefunctions for different types of atoms are stored separately without additional computation.

\subsection{Algorithms implemented in KSSOLV 2.0 of conventional calculations}

In this section, we describe several standard algorithms implemented in KSSOLV 2.0 for conventional calculations, These include the self consistent field (SCF) iteration and direct energy minimization, matrix diagonalization and geometry optimization. In addition, we describe a method called SCDM (Select Column of the Density Matrix) used to perform orbital localization.

\subsubsection{Self consistent field iteration and direct minimization}\label{sec:scfdcm}

When LDA or GGA is used in $\Vxc$, the KS eigenvalue problem can be formulated as a set of nonlinear
equations satisfied by the ground state electron density or
potential~\cite{HohenbergKohn1964}, i.e.,
\begin{equation}
\rho = f_{\mathrm{KS}}(\rho),
\label{eq:ksmap}
\end{equation}
where $f_{\mathrm{KS}}(\cdot)$ is known as the Kohn-Sham
map~\cite{kohn1965}. This formulation suggests that the KS equations can
be solved by a quasi-Newton method in which the Jacobian of the Kohn-Sham
map is approximated. To be specific, the approximation to $\rho$ can be
updated as
\begin{equation}
\rho^{k+1} = \rho^{(k)}-\hat{J}(\rho^{(k)})\left[\rho^{(k)}-f_{\mathrm{KS}}(\rho^{(k)})\right],
\label{eq:qnewton_rho}
\end{equation}
where $\hat{J}$ is an approximate Jacobian. This approach is generally
known as the self-consistent field (SCF) iteration in the physics
literature and is implemented in KSSOLV. In this approach, the evaluation
of the Kohn-Sham map on the right hand side of \eqref{eq:qnewton_rho}
requires computing eigenvalues and eigenvectors of the Kohn-Sham
Hamiltonian defined at $\rho^{(k)}$, which will be discussed in the next
section. 

There are many ways to approximate the Jacobian of the Kohn-Sham map. The
simplest is to take $\hat{J} = \beta I$, where $0<\beta<1$ is a small
constant and $I$ is the identity matrix. Such an approximation yields the
so-call simple mixing scheme described by
\begin{equation}
\rho^{(k+1)} = \beta \rho^{(k)}+(1-\beta)f_{\mathrm{KS}}(\rho^{(k)}).
\end{equation}

More sophisticated Jacobian approximate schemes include the
Anderson~\cite{anderson} or Pulay~\cite{pulay} mixing, two types of
Brodyen's method and Kerker mixing~\cite{kerker}, which can also be viewed as a way to
accelerate the convergence of the quasi-Newton iteration
\eqref{eq:qnewton_rho} by preconditioning the nonlinear equation
\eqref{eq:ksmap}~\cite{precondition}. All these Jacobian approximation and
precondition methods have been implemented in KSSOLV 2.0. In earlier
work~\cite{preconditioning}, we have demonstrated how new preconditioners
can be easily implemented in KSSOLV.

An alternative approach to solving the Kohn-Sham problem is to solve the constrained minimization problem \eqref{eq:etotmin} directly. This approach is known as direct minimization. In KSSOLV, we implement a direct constrained minimization algorithm presented in~\cite{dcm1}. In each step of the algorithm, a subspace that consists of the current approximation to the Kohn-Sham orbitals, the preconditioned gradient of the Lagrangian and previous search direction is constructed. The update of the wavefunction approximation is obtained by minimizing the total energy \eqref{eq:energy} within this subspace.  Trust region techniques~\cite{dcm2} can be used in the DCM algorithm to stabilize the convergence of the iterative minimization procedure. This is particularly useful for metallic systems at low temperature.

\subsubsection{Eigensolver}\label{sec:eigen}
When the SCF iteration is used to solve the Kohn-Sham problem, the most time-consuming part of the computation is the evaluation of the Kohn-Sham map $f_{\rm{KS}}(\rho)$. At a finite temperature, the Kohn-Sham map is defined as
\begin{equation}
f_{\rm{KS}}(\rho) = \mbox{diag} \left[ \left(I+e^{\frac{H(\rho)-\mu I}{\kappa_B T}}\right)^{-1} \right],
\label{eq:fks}
\end{equation}
where $\mu$ is the chemical potential, $\kappa_B$ is the Boltzmann factor and $T$ is the temperature. The matrix exponential in \eqref{eq:fks} can be evaluated by a partial  spectral decomposition of $H$.  In the limit of $T=0$, \eqref{eq:fks} reduces to \eqref{eq:rho}. In this case, we only need to compute the leftmost $N_e$ eigenvalues of $H(\rho)$ and their corresponding eigenvectors. For a finite temperature calculation, we need to compute a few extra eigenvalues $\varepsilon_j$ and eigenvectors $\psi_j$ that have non-negligible occupation numbers $1/(1+\exp({\frac{\varepsilon_j-\mu}{\kappa_B T}}))$.

Because $H$ is not explicitly stored as a matrix in KSSOLV, iterative eigensolvers are appropriate for computing the desired eigenvalues and eigenvectors of $H$.  In KSSOLV, the default eigensolver employed in a SCF iteration uses the locally optimal block preconditioned conjugate gradient (LOBPCG) method~\cite{duersch2018robust}.  The method can be viewed as a constrained minimization method for solving the equivalent trace minimization problem
\begin{equation}
\min_{X^TX=I}\mathrm{trace}(X^T H X),
\label{eq:tracemin}
\end{equation}
where $X$ is a matrix that contains the discretized Kohn-Sham orbitals.  Similar to other plane-wave based Kohn-Sham solvers, KSSOLV stores the plane-wave expansion coefficients of each Kohn-Sham orbitals in $X$. In each LOBPCG iteration, we need to multiply $H$ with a set of vectors. This is done by multiplying the kinetic energy operator $\hat{T}$, the nonlocal part of the ionic pseudopotential operator $\Vion$ and the Hartree potential operator $\Vhartree$ with the plane-wave expansion coefficients in the reciprocal space and transforming the result to a real space grid (via FFTs) on which the local part of $\Vion$ and the local exchange-correlation potential operator $\Vxc$ are applied.

In addition to LOBPCG, KSSOLV 2.0 also includes an implementation of the Davidson-Liu~\cite{davidson197514} algorithm. The algorithm can be viewed as a generalization of the LOBPCG method in the sense that the update of the eigenvector is obtained by projecting the $\hat{H}$ into a progressively larger subspace constructed from the juxtaposition of preconditioned gradients of the Lagrangian and the subspace constructed in the previous iteration, and solving the projected eigenvalue problem. When the dimension of the subspace
reach a prescribed limit, the procedure is restarted with the most recent approximation of the eigenvectors. Clearly, there is a trade-off between the per iteration cost of the Davidson method, which becomes higher if the maximum allowed dimension of the subspace ($m_D$) is large, and the number of restarts required to reach convergence, which becomes lower when $m_D$ is larger.

Both the LOBPCG and Davidson solvers are block eigensolver, i.e., in each iteration, the Hamiltonian is applied to a block of vectors, and many other linear algebra operations in the solver can be expressed in terms of level-3 BLAS operations. These features can significantly enhance the concurrency of the computation and take advantage of parallel computer architecture and memory hierarchy. 

Another block algorithm that has been demonstrated to be very efficient for solving large-scale Kohn-Sham eigenvalue problem is the Chebyshev filter subspace iteration (CheFSI)~\cite{chefsi}. This method is implemented in KSSOLV 2.0 also. The CheFSI method constructs a properly shift and scaled $m$th degree Chebyshev polynomial $T_m$ to amplify the contribution of the desired eigenvectors when $T_m(H)$ is applied to a set of properly prepared vectors in a subspace iteration.  The multiplication of $T_m(H)$ with a block of vectors $X$ can be implemented via a 3-term recurrence. We do not need to solve a projected eigenvalue problem, i.e., we do not need to perform the Rayleigh-Ritz procedure in each subspace iteration. This can significantly reduce the computational cost for the problem with a large number of electrons. The orthonormality of $X$ in each iteration can be maintained by using the Cholesky QR procedure, which is generally efficient. The Rayleigh-Ritz procedure only needs to be performed at the end of subspace iteration to compute the occupation number for each desired eigenvalue.

Two other methods that are designed to reduce the cost of Rayleigh-Ritz calculations in an eigensolver are the project preconditioned conjugated gradient (PPCG)~\cite{ppcg} method and the residual minimization method with direct inversion in iterative subspace (RMM-DIIS)~\cite{rmmdiis} acceleration. Both methods are implemented in KSSOLV 2.0 also. Although for most small to medium sized problems to be solved by KSSOLV, the Rayleigh-Ritz cost in the block algorithms discussed above is relatively small, the availability of additional eigensolvers allows us to test and compare convergence properties of these algorithms.

In PPCG, we apply the LOBPCG algorithm to each approximate eigenvector separately, i.e. running the unblocked version of the LOBPCG method for each desired eigenpair for a fixed number of iterations. The Rayleigh-Ritz procedures in these runs only need to solve a set of $3 \times 3$ projected eigenvalue problems.  A global Rayleigh-Ritz procedure for all desired eigenpairs is only applied periodically at the end of a fixed number of unblocked LOBPCG iterations.

In RMM-DIIS, each approximate eigenpair is updated separately by minimizing the residual (instead of the Rayleigh-quotient) associated with the approximate eigenpair within a progressively larger subspace incrementally constructed from a set of previously approximations to the same eigenpair.  When the initial guess of the desired eigenpairs are sufficiently accurate, no Rayleigh-Ritz procedure is ever needed in RMM-DIIS. This feature of the algorithm makes it ideal for solving large-scale problems that contain many electrons on a parallel computer on which the refinement of each eigenpair can be carried out independently.

One of the key features of the eigenvalue problems solved in each SCF iteration is that the accuracy required for the desired eigenpairs is generally lower in early SCF iterations and higher in later iterations when self-consistency is nearly reached.  This is due to the fact that in early iterations of the quasi-Newton algorithm used to solve \eqref{eq:ksmap}, the residual term $\rho^{(k)} - f_{\mathrm{KS}}(\rho^{(k)})$ on the right-hand side of \eqref{eq:qnewton_rho} is relatively large even if $f_{\mathrm{KS}(\rho^{(k)})}$ is evaluated to full accuracy.  As a result, we may lower the accuracy requirement for $f_{\mathrm{KS}}(\rho^{(k)})$, and consequently the accuracy requirement for the solution of the eigenvalue problem could be reached.  As the SCF iteration converges, a more accurate evaluation  of $f_{\mathrm{KS}}(\rho^{(k)})$. Therefore, in KSSOLV 2.0, we use an adaptive strategy to define the convergence criterion for each eigenpair.  An approximate eigenpair $(\theta, \psi)$ is considered converged if the relative residual norm
\begin{equation*}
\|r\|/|\theta| = \| H\psi - \theta \psi \| / |\theta|,
\end{equation*}
is less than a tolerance $\tau^{(k)}$, where 
\begin{equation}
\tau = \min (\tau_0, \|\rho^{(k)} - f_{\mathrm{KS}}(\rho^{(k)})\|/\|\rho^{(k)}\|),
\end{equation}
and $\tau_0$ is a maximal error tolerance set to $10^{-2}$ by default, but can be changed by a user.

Moreover, at a finite temperature, lower accuracy can be tolerated for partially occupied states with low occupation numbers.

\subsubsection{Geometry optimization}\label{sec:relaxatom}
In KSSOLV 2.0, we include the functionality to compute atomic forces
which are the derivatives of $\Etot$ defined in \eqref{eq:energy} with respect to atomic coordinates.
The derivatives can be taken with respect to either the cartesian coordinates or relative coordinates of the atoms~\cite{cpmd}. The calculations of these forces make use of the Hellmann-Feynman theorem~\cite{feynman1939forces}. The availability of atomic forces allows us to optimize the atomic structure of a molecule or solid. The calculations are often referred to geometry optimization or structure relaxation.
The goal of the optimization is to minimize the total energy of the atomistic system with respect to atomic coordinates. The atomic forces simply yield the gradient of the objective function.

KSSOLV 2.0 leverages the standard unconstrained minimization algorithm implemented in MATLAB's optimization box.  The user has the option of using MATLAB's {\tt fminunc}(Find minimum of unconstrained
multivariable function) function to perform the optimization.  By default, {\tt fminunc} uses the BFGS (Broyden–Fletcher–Goldfarb–Shanno)~\cite{head1985broyden} quasi-Newton algorithm which constructs approximations to the Hessian of the energy using gradients computed in successive quasi-Newton iterations.  One can also choose the {\tt trust region} algorithm, which is based on the interior-reflective Newton method described in~\cite{irnewton}.

In addition to algorithms implemented in MATLAB Optimization toolbox, one can also use other algorithms such as the limited memory BFGS~\cite{lbfgs} algorithm implemented in the HANSO package~\cite{hanso} or the nonlinear conjugate algorithm~\cite{nlcg} implemented by Overton~\cite{nlcgpack}.  Both algorithms contain a number of parameters that a user can experiment with and adjust.
KSSOLV 2.0 also provides ample flexibilities to utilize other optimization algorithms. For example, we also implemented a version of the FIRE (Fast Inertial Relaxation Engine)~\cite{FIRE} algorithm.

\subsubsection{Orbital localization via selected column of density matrix}\label{sec:scdm}
It is well known that electrons in insulating systems obey the  nearsightedness principle, i.e. local electron properties such as the electron density $\rho(\br)$ only depend significantly on the effective potential at nearby points. 
Mathematically, the nearsightedness principle translates into the decay property of 
the single-particle density matrix associated with the ground state of the 
atomistic system, i.e., the magnitude of the matrix elements of the density matrix decays rapidly 
away from the diagonal.  A direct consequence is that the occupied 
Kohn-Sham orbitals can be rotated to a set of functions that span the same invariant space, but have approximately localized support. 
These localized orbitals can be used to develop linear scaling methods for solving the Kohn-Sham problem and to develop efficient post-DFT methods~\cite{localorbital,MarzariMostofiYatesEtAl2012}.

There are several ways to construct localized orbitals. One of the most known technique is the
maximally localized Wannier functions (MLWFs) proposed by Marzari and Vanderbilt~\cite{mlwfs}. The 
MLWF method requires solving a nonlinear optimization problem, whose results can sometimes depend sensitively to the initial guesses. Recently, an alternative method called Selected Column of the Density Matrix (SCDM)~\cite{scdm} 
has been proposed to construct localized orbital using a simple linear algebraic procedure.
Suppose $\Psi$ is an $N \times N_e$ matrix containing $N_e$ approximate Kohn-Sham orbitals
on $N$ real space grid points. The SCDM method performs a rank revealing QR factorization of 
$\Psi^*$ first to yield
\begin{equation}
\Psi^{*}\Pi = Q R,
\label{eq:rrqr}
\end{equation}
where $\Pi$ is a column permutation matrix that moves maximally linearly independent columns of
$\Psi^*$ (or a row permutation matrix that moves maximally linear independent rows of $\Psi$) to the leading column (row) positions, $Q$ is a $N_e\times N_e$ unitary matrix and $R$ is $N_e \times N$ matrix with the leading $N_e$ columns being a upper triangular matrix. The magnitudes of the diagonal matrix elements of the leading columns of $R$ are in a decreasing order.  

Localized orbitals can be computed simply by performing a matrix multiplication
\begin{equation}
\Phi = \Psi \Psi_C^*,
\end{equation}
where $\Psi_C$ represents the leading $N_e$ rows of the row permuted $\Psi$ where the permutation is defined by the permutation matrix $\Pi$ obtained in \eqref{eq:rrqr}. Note that the localized columns in
$\Phi$ are not necessarily orthonormal. To obtain an orthonormal set of orbitals $\tilde{\Phi}$ that 
remain to be localized, we simply perform a Cholesky factorization of the matrix 
$P_{C,C} = \Psi_C \Psi_C^*$, i.e., 
\begin{equation}
P_{C,C} = LL^*,
\label{eq:Lfact}
\end{equation}
and solve the following set of linear equations using the Cholesky factor $L$ obtained in \eqref{eq:Lfact}
\[
\tilde{\Phi} L^* = \Phi.
\]

The SCDM method has been implemented in KSSOLV. We refer readers to~\cite{scdm} for the theoretical justification of this method and how localized orbitals constructed from the SCDM procedure can be used to speedup the Hartree-Fock exchange energy calculation~\cite{scdm-a1,gace}.

\subsection{Accelerated algorithms implemented in KSSOLV 2.0 for hybrid functional DFT calculations }\label{sec:C}

In KSSOLV 2.0, we implement several new algorithms to accelerate hybrid functional DFT calculations. 
The main challenge in performing a hybrid functional DFT calculation is the efficient treatment of the (screened) Hartree-Fock exchange potential defined as 
\begin{equation}
\Vhse(\br,\brp) = -\sum_{j=1}^{N_e} \psi_j(\br) \psi_j^\ast(\brp) K(\br,\brp),
\label{eq:Vhse}
\end{equation}
where $K(\br,\brp)$ is either the Coulomb kernel $1/|\br -\br'|$ or a screened Coulomb kernel of the form $\mathrm{efrc}(\mu |\br - \brp|)/|\br - \brp|$.  This non-local potential is part of the exchange-correlation potential $\Vxc$ in a hybrid functional DFT Hamiltonian.

In KSSOLV 2.0, we do not explicitly construct $\Vhse$, which is a dense matrix, in either the real or reciprocal space. The $\Vhse$ operator is applied to a set of wavefunctions $\{\varphi_i\}$ as follows
\begin{equation}
\Vhse \varphi_i=-\sum_{j=1}^{N_e} \psi_j(\br) \int \varphi_{i}\left(\br^{\prime}\right) \psi_{j}^{*}\left(\br^{\prime}\right)
K(\br,\brp) d \brp. 
\label{eq:applyVhse}
\end{equation}
The evaluation of the integral on the right hand side of \eqref{eq:applyVhse} requires solving a set of Poisson equations. This can be done by using FFT based convolution. However, because the summation in \eqref{eq:applyVhse} is over $N_e$ terms, we need to solve $\mathcal{O}(N_e^2)$ Poisson equations in total per iteration in an iterative eigensolver used to compute the lowest $N_e$ eigenpairs of the hybrid functional Hamiltonian. The excessive number of FFTs used to solve many Poisson equations is the reason that hybrid functional DFT calculation is orders of magnitude more expensive than LDA or GGA DFT calculations in other plane-wave DFT software tools. KSSOLV 2.0 uses several recently developed algorithms to reduce the complexity of hybrid functional DFT calculation. These algorithms include 1) the interpolative separable density fitting (ISDF) method for reducing the number of Poisson equations to be solved; 2) the use of inner and outer iterative schemes in combination with the adaptive compressive exchange (ACE) operator method to further reduce the total number of Poisson equations to be solved; 3) a special projector commutator direct inversion of iterative subspace (PC-DIIS) method for accelerating the outer SCF iteration.  We will briefly describe each one of these algorithms below.

\subsubsection{ISDF (interpolative separable density fitting decomposition)}\label{isdf}
If we place the right hand sides of the Poisson equations to be solved in \eqref{eq:applyVhse} for $i=1,2,...,N_e$ in a matrix $Z$, defined as  
\begin{equation}
Z=\{\varphi_i(\br) \psi^\ast_j(
\br) \}, \ \ i,j = 1,2,...,N_e,
\label{eq:zmat}
\end{equation}
we can see that the rank $Z$ is less than $N_e^2$ 
if $\varphi_i(\br)$ and $\psi^\ast_j(\br)$ are discretized on a real space grid with $N_g = \mathcal{O}(N_e)$ grid points, which is the case for systems that are sufficiently large.  As a result, we can rewrite $Z$ 
as 
\begin{equation}
Z=\Theta C,
\label{eq:zfactor}
\end{equation}
where $\Theta$ is $N_g \times N_\mu$ and 
$C$ is $N_\mu \times N_e^2$ and $N_\mu = \mathcal{O}(N_e)$. Columns of $\Theta$ can be viewed as a set of numerical auxiliary basis $\{\zeta_{\mu}(\br)\}$, $\mu = 1,2,...,N_\mu$ that span the same space defined by the pair product basis $\{\varphi_i(\br)\psi^\ast_j(\br)\}$.  Consequently, we can evaluate \eqref{eq:applyVhse} by first computing
\begin{equation}
V_\mu^{\zeta} = \int K(\br,\br') \zeta_\mu(\brp) d \brp,
\label{eq:Vmu}
\end{equation}
for all $\mu=1,2,...,N_\mu$, which requires solving $N_\mu$ Poisson equations.  If we use $V^\zeta$ to denote the matrix that contains $V_\mu^\zeta$'s as its columns, \eqref{eq:applyVhse} can be then evaluated as 
\begin{equation}
\Vhse \varphi_i = -\sum_j \psi_j(\br) V^\zeta c_{ij},
\end{equation}
where $c_{ij}$ is the column of $C$ indexed by $i$ and $j$ consistent with the column indexing scheme used in \eqref{eq:zmat}.

Although the computational procedure for evaluating \eqref{eq:Vhse} now requires solving only $N_\mu = \mathcal{O}(N_e)$ Poisson equations, the overall complexity of the algorithm hinges on an efficient factorization of $Z$ in \eqref{eq:zfactor}.  From an accuracy standpoint, the optimal factorization can be obtained by performing a singular value decomposition (SVD) of $Z$. However, such a factorization is costly.  

In~\cite{hu2017interpolative}, the ISDF technique is used to obtain an approximate factorization that is much more efficient and sufficiently accurate.  In ISDF, each entry of the $C$ matrix is chosen to be $\varphi_i(\br_\mu)\psi^\ast_j(\br_\mu)$ for a set of carefully chosen real space grid points $\br_\mu$.  The auxiliary basis vectors in $\Theta$ can be obtained by solving a linear least square problem. Due to the separable nature of the pair product basis in $C$, this least square problem can be solved efficiently.  We will refer readers to~\cite{hu2017interpolative,lee2019systematically} for computational details of the ISDF method.  We should note that in this approach $N_\mu$ is a parameter that a user needs to choose in advance. Typically, $N_\mu$ is a small multiple of $N_e$, e.g. $2N_e$. As a result, the use of ISDF allows us to reduce the overall computational complexity of $\Vhse$ related operation to $O(N_e^3)$.

\subsubsection{ACE (adaptively compressed exchange)}\label{ace}
Due to the high cost associated with the application of the Hartree-Fock exchange operator $\Vhse$ to a set of wavefunctions, the iterative solution of the Kohn-Sham problem for hybrid functional DFT is separated into inner and outer SCF iterations. At the beginning of each outer SCF iteration, $\Vhse$ is updated with the most recent approximations to the Kohn-Sham orbitals $\{\psi_{j}\}$.  This $\Vhse$ is then fixed in the corresponding inner SCF iterations in which only the charge density $\rho$ and potential terms that depends on $\rho$ are updated.

However, as we can see in \eqref{eq:applyVhse}, even when $\{\psi_j\}$ and $\Vhse(\{\psi_j\})$ are fixed, applying $\Vhse(\{\psi_j\})$ to a set of orbitals $\varphi_i$, $i = 1,2,...,N_e$ is costly. Although we can use ISDF to reduce the number of Poisson solves from $\mathcal{O}(N_e^2)$ to $\mathcal{O}(N_e)$, performing the ISDF procedure and solving $\mathcal{O}(N_e)$ Poisson equations in each inner SCF iteration is still quite costly.

To reduce the computational cost of each inner iteration, 
Lin~\cite{JCTC_12_2242_2016_ACE} proposed the construction of an approximate $\Vhse$ using a procedure called the Adaptively Compressed Exchange Operator (ACE) algorithm.  The approximate $\Vhse$, denoted by $\Vace$, is constructed to satisfy the condition
\begin{equation}
\Vhse \psi_j = \Vace \psi_j,
\end{equation}
where $\psi_j$, $j = 1,2,...,N_e$ is the set of approximate Kohn-Sham orbitals available at the beginning of each outer SCF iteration,  The ACE construction yields a low-rank operator of the form
\begin{equation}
\Vace = - \sum_{i,j=1}^{N_e} W_i(\br) B_{ij} W_j^\ast(\brp),
\label{eq:Vace}    
\end{equation}
where 
\begin{equation}
W_{i}(\mathbf{r}) =\left(\Vhse[\{\psi_i\}] \psi_{i}\right)(\mathbf{r}), \quad i=1, \ldots, N_{\mathrm{e}},
\label{eq:Wace}
\end{equation}
$B =M^{-1}$ and the $(k,l)$th element of the overlap matrix $M$ is $M_{k l} =\int \psi_{k}(\mathbf{r}) W_{l}(\mathbf{r}) \mathrm{d} \mathbf{r}$.

By constructing an ACE approximation of $\Vhse$ in the low rank form \eqref{eq:Vace}, we can apply $\Vace$ to a set of orbitals $\{\varphi_i\}$ in each SCF inner iteration by using two matrix-matrix multiplications.  This type of BLAS3 dense linear algebra operations are extremely efficient on modern high performance computers. 

We should note that the construction of $\Vace$ in each outer SCF iteration requires solving $\mathcal{O}(N_e^2)$ Poisson equations in \eqref{eq:Wace} just as $\mathcal{O}(N_e^2)$ Poisson equations need to be solved in \eqref{eq:applyVhse}.  The number of Poisson equations to be solved can be reduced to $\mathcal{O}(N_e)$ by using the ISDF technique discussed above.  Therefore, by combining ACE with ISDF, we can significantly reduce the complexity of hybrid functional DFT calculation as reported in~\cite{aceisdf}.

\subsubsection{PC-DIIS (projected commutator direct inversion in the iterative subspace)}\label{pcdiis}
As we indicated in section~\ref{sec:scfdcm}, when LDA and GGA are used in $\Vxc$,  the KS eigenvalue problem can be formulated as a set of nonlinear
equations \eqref{eq:ksmap} satisfied by the ground state electron density $\rho$.  For hybrid functional DFT, a similar nonlinear equation should be defined in terms of the density matrix $P = \sum_{j=1} \psi_j(\br)\psi_j^\ast(\brp)$ at zero temperature.
Alternatively, one can define a nonlinear equation in terms of the commutator between $H(P)$ and $P$. Upon convergence, $P$ satisfies
\begin{equation}
H(P)P-PH(P) = 0.
\label{eq:commhp}
\end{equation}

The outer SCF iteration used to solve the hybrid functional KS-DFT problem can be viewed as a quasi-Newton method for finding the solution of \eqref{eq:commhp}.  When the density matrix can be formed explicitly, one can use the direct inversion of iterative subspace (DIIS) method proposed by Pulay~\cite{pulay} to solve  \eqref{eq:commhp}. This is the approach often used to solve the Hartree-Fock equation in quantum chemistry.  Given a few previous approximations to the density matrix $P^{(i-1)}$, $P^{(i-2)}$,...,$P^{(i-\ell)}$, for some constant $\ell < i$, the DIIS method or commutator DIIS (C-DIIS) method constructs a new approximation to the density matrix in the $i$th iteration as
\begin{equation}
\tilde{P} = \sum_{k=1}^{\ell} \alpha_k P^{(i-k)},
\label{eq:combineP}
\end{equation}
where the coefficients $\alpha_k$ are chosen to solve the following constrained minimization problem
\begin{equation}
\min_{\sum_k \alpha_k = 1}
\| \alpha_k R[P^{(i-k)}] \|_F,
\label{eq:mindiis}
\end{equation}
where 
\begin{equation}
R[P^{(i-k)}] =  H[P^{(i-k)}] P^{(i-k)} - P^{(i-k)} H[P^{(i-k)}]
\end{equation}
and $\|\cdot \|_F$ is the Frobenius norm.

When the Kohn-Sham orbitals $\psi_j$'s are discretized by plane-wave expansions, it is generally not practical to construct the density matrix $P$ explicitly and solve the minimization problem \eqref{eq:mindiis} directly because the density matrix dimension is so large($N_{g} \times N_{g}$) within a plane-wave basis set. In~\cite{hu2017projected}, a projected commutator DIIS (PC-DIIS) method was proposed to solve a projected minimization problem in which the matrix $R[P^{(i-k)}]$ in \eqref{eq:mindiis} is replaced by
\begin{equation}
R[P^{(i-k)}]\Phiref = H\Psi^{(i-k)} S^{(i-k)} - \Psi^{(i-k)}T^{(i-k)},
\label{eq:rpsi}
\end{equation}
where $\Phiref$ is a set of reference orbitals to be defined later and $\Psi^{(i-k)}$ is a matrix that contains approximate Kohn-Sham orbitals $\psi_j$'s obtained in the $(i-k)$th outer SCF iteration,
$S^{(i-k)} =\langle \Psi^{(i-k)}, \Phiref \rangle$ and $T^{(i-k)} = \langle H\Psi^{(i-k)}, \Phiref \rangle$. Note that we dropped the density matrix $P^{(i-k)}$ in the Hamiltonian $H$ above to simplify the notation. The projected residual \eqref{eq:rpsi} can be computed without forming $P^{(i-k)}$ explicitly.
We will refer readers to~\cite{hu2017projected} for the theoretical justification for using \eqref{eq:rpsi} in the objective function of the minimization problem \eqref{eq:mindiis}.  The solution of the alternative minimization problem is used to construct an intermediate set new approximation to Kohn-Sham orbitals as
\[
\tilde{\Psi} = \sum_{k=1}^{\ell} \alpha_k \Psi^{(i-k)}.
\]

The eigenvectors of $H[\tilde{\Psi}]$ then form the approximate Kohn-Sham orbitals $\Psi^{(i)}$ in the $i$th SCF iteration. Self-consistency is achieved when the norm of  $H[\Psi^{(i)}]\Psi^{(i)}-\Psi^{(i)}\Lambda^{(i)}$ is sufficiently small, where $\Lambda^{(i)}$ is a diagonal matrix containing the corresponding eigenvalues of $H[\tilde{\Psi}]$.

We should note that the reference orbitals in $\Phiref$ can be chosen to be any linearly independent functions that approximates the desired Kohn-Sham orbitals. They can be chosen as a set of Kohn-Sham orbitals obtained in an LDA or GGA calculation.  Also, the constrained minimization problem \eqref{eq:mindiis} can be easily converted to an unconstrained least square minimization problem by substituting $\alpha_1$ in the objective function with $1-\sum_{k=2}^{\ell} \alpha_k$. We will refer readers to~\cite{hu2017projected} for algorithmic and computational details.

\section{Object Oriented Design}
\label{sec:design}

Object-oriented programming (OOP) is a modern design paradigm developed to define
data and functions together as an object. \kssolv{} adopts OOP features
in \matlab{} and implements many key quantities required in the numerical solution of \eqref{eq:kseig} as classes. In
\kssolv{}, there are several basic classes and some more advanced classes.
The basic classes include the {\tt Atom}, {\tt Molecule}, {\tt Crystal}, {\tt PpData}, {\tt PpVariable}, {\tt Ggrid}, and {\tt IterInfo} classes.  The {\tt Atom}, {\tt Molecule} and {\tt Crystal} classes are created to represent and encapsulate all relevant properties of an atom, a molecule and a crystal respectively. All relevant features of an
object, \eg, the mass of an atom, the positions of all atoms within a molecule, the energy cut-off used for plane-wave expansion is kept as attributes (member variable) of the object. 
Since a crystal shares many features with a molecule,
the {\tt Crystal} class is defined as a derived class of the {\tt Molecule} class with additional attributes such as the positions of k-point samples and their corresponding weights.  The {\tt PpData} and {\tt PpVariable} classes are two
classes that encapsulate a variety of information related to the pseudopotential. The {\tt PpData} class is used to encapsulate the raw data read from a pseudopotential file, and the {\tt PpVariable} class stores the actual pseudopotentials associated with all atomic species contained in a molecule (or crystal). 
The {\tt Ggrid} class is used to provide a compact representation of 
reciprocal space grid points enclosed within a sphere of a fixed radius determined by the kinetic energy cutoff $E_{\rm{cut}}$. Note that in a plane-wave based DFT calculation, the plane-wave coefficients associated with reciprocal grid points outside of this sphere are set to zero, and thus not stored. Finally,
the {\tt IterInfo} class is a bookkeeping class used to simply record information related to the SCF/DCM iterations.
All these basic classes are designed as data containers to simplify the interfaces in \kssolv{}. The member functions in these classes are used to process data within the class but do not interfere with data outside the class.

In the following, we will introduce advanced classes in \kssolv{} one by
one in detail, \ie, the {\tt Wavefun}, {\tt Ham}, {\tt BlochWavefun} and {\tt BlochHam} classes.

\subsection{The Kohn-Sham wavefunction class}

Kohn-Sham orbitals are key quantities used and updated throughout a KS-DFT calculation. We created a class, called {\tt Wavefun}, to encapsulate all relevant information contained in these orbitals. This class contains matrix attributes that keep either the values of wavefunctions on a real space grid or plane-wave expansion coefficients on a compressed reciprocal space grid. Standard algebraic operations applied to a {\tt Wavefun} object are overloaded.
They include element-wise operations such as the absolute value, the addition, subtraction, pointwise multiplication and pointwise powering. These operations typically return 
a {\tt Wavefun} object. Other operations such as the matrix norm and the inner product of two sets of wavefunctions return a scalar or a matrix.
Other commonly used operations such as the QR factorization, SVD and 3D (inverse) Fast Fourier Transform are overloaded as well. Listing~\ref{code:oop-wavefun} provides some simple examples of overloaded operations on a {\tt Wavefun} object.

\begin{lstlisting}[language=Matlab, caption=Setting up a {\tt Wavefun} 
object~\label{code:oop-wavefun}]
% X is a @Wavefun object of size N by k
X = Wavefun(...);

% Q is a @Wavefun of the same size as X,
% and R is a k by k upper triangular matrix
[Q,R] = qr(X,0);

% U is a @Wavefun of the same size as X,
% and S and V are k by k matrices
[U,S,V] = svd(X,0);

% Y is a @Wavefun object of size N by p
% on the same grid points as X
Y = Wavefun(...);

% Z is a @Wavefun object of size N by (k+p)
Z = [X Y];

% T is a submatrix of Z with its odd row index
% and last k column index.
% T is a @Wavefun object of size N/2 by k
% Then the corresponding submatrix of Z is reset
% to a random matrix.
T = Z(1:2:end,end-k+1:end);
Z(1:2:end,end-k+1:end) = randn(N/2,k);
\end{lstlisting}

Other overloaded functions include the concatenation of {\tt Wavefun} objects, the selection of one or a subset of wavefunctions, which are unique in MATLAB. We allow a {\tt Wavefun} object to be multiplied with a matrix also when the dimension of the matrix contained in the {\tt Wavefun} object is compatible with that of second matrix to be multiplied.


\subsection{The Hamiltonian class}\label{sec:hamclass}

Even though the Kohn-Sham Hamiltonian is not stored as a matrix in KSSOLV, it is convenient to create a {\tt Ham} class that allows us to easily apply the Hamiltonian to a {\tt Wavefun} object. 
The {\tt Ham} class encapsulates the kinetic and potential energy components of the Hamiltonian in either real space or reciprocal space representation as well as the charge density associated with the Hamiltonian. Member functions are created to make it easy to update the Hamiltonian when the charge density is changed.
The multiplication of a {\tt Ham} object {\tt H} and a {\tt Wavefun} object {\tt X} can be simply performed as {\tt H*X} with all the details resulting from the conversion from the real space to the reciprocal space and back to the real space representation of the wavefunction hidden from the user. See Listing~\ref{code:oop-ham} for how a {\tt Ham} object is created and used.  Furthermore, we have also implemented several functions such as the MINRES and GMRES functions for solving the linear system of involving a shifted Kohn-Sham Hamiltonian operator. 
\begin{lstlisting}[language=Matlab, caption=Setting up a Hamiltonian
object~\label{code:oop-ham}]
% H is a @Ham object of size N by N
H = Ham(...);

% X is a @Wavefun object of size N by k, and Y is
% a @Wavefun of the same size
X = Wavefun(...);
Y = H*X;

% V is a random matrix of size N by k, and U is
% a matrix of the same size
V = randn(N,k);
U = H*V;

% BH is a @BlochHam object of size N by N for
% m k-points
BH = BlochHam(...);

% BX and BY are a @BlochWavefun objects of size
% N by k for m k-points
BX = BlochWavefun(...);
BY = BlochWavefun(...);

% BH is applied to BX and saved at BY
for i = 1:m
	BY{i} = BH{i}*BX{i};
end
\end{lstlisting}

\subsection{Wavefunction and Hamiltonian classes for solids}

For periodic systems, we have created the {\tt BlochWavefun} and
{\tt BlockHam} classes to encapsulate data elements required to
represent Bloch wavefunctions and Hamiltonian. These classes allow
users to specify a k-point sampling and the associated weights. They
are containers of the {\tt Wavefun} and {\tt Ham} type variables
respectively.  Listing~\ref{code:oop-ham} contains an example of how
these two classes are used.

\section{Use Cases}
\label{sec:usecase}
In this section, we will illustrate some key features of KSSOLV through some use cases.  The main workflow for using KSSOLV to perform an electronic structure calculation of a molecule or solid involves
\begin{enumerate}
    \item Setting up the system;
    \item Calling an appropriate function to solve the Kohn-Sham problem or perform a geometry optimization;
    \item Examining, post-processing and visualizing the results.
\end{enumerate}

We will use a simple example to demonstrate how to perform a basic calculation in section~\ref{setup}.  One of the key advantages of KSSOLV is that it allows users to try different algorithms and algorithmic parameters. We will illustrate how this can be achieved in KSSOLV by properly setting different options and comparing results. The object-oriented design of KSSOLV enables developers to prototype and implement new algorithms with ease. We will give an example to show some of the key features that make prototyping new algorithms easy in KSSOLV.  Finally, the MATLAB performance profiler allows developers to identify the main computational bottleneck of the calculation and develop strategies to improve computational efficiency.

\subsection{Setting up and solving a simple problem}\label{setup}
Before we perform an electronic structure calculation for a molecule or a solid, we must first set up the system.  This step entails selecting the constituent atoms and defining their atomic coordinates. In addition, we must define a sufficiently large unit (super)cell that contains all constituent atoms. The list of atoms and their coordinates as well as the supercell are used to define a {\tt Molecule} object.  
For example, in Listing~\ref{code:setup}, we show how a silane molecule ($\mathrm{SiH}_4$) is set up. 
\begin{lstlisting}[language=Matlab, caption=Setup A System~\label{code:setup}]
kssolvpptype('ONCV_PBE-1.0', 'UPF');
%kssolvpptype('pz-hgh', 'UPF');
%
% 1. construct atoms
%
a1 = Atom('Si');
a2 = Atom('H');
atomlist = [a1 a2 a2 a2 a2];
%
% 2. set up a supercell
%
C = 10*eye(3);
%
% 3. define the coordinates the atoms
%
redxyz = [
 0.0     0.0      0.0
 0.161   0.161    0.161
-0.161  -0.161    0.161
 0.161  -0.161   -0.161
-0.161   0.161   -0.161
];
xyzlist = redxyz*C';
%
% 4. Configure the molecule (crystal)
%
mol = Molecule('supercell',C,'atomlist',
      atomlist,'xyzlist',xyzlist, ...
      'ecut',12.5,'name','SiH4' );
\end{lstlisting}

In this script, which can be found in the {\tt kssolv2.0/examples} directory, we first choose the pseudopotential type by using {\tt kssolvpptype}. The Optimized Norm-Conserving Vanderbilt (ONCV)~\cite{oncv} is chosen (by default). Changing it to another type of NCPP, e.g. the Hartwigsen-Goedecker-Hutter (HGH)~\cite{hgh} pseudopotential simply involves uncommenting the second line of the code snippet.  In \kssolv{} 2.0, users can adopt NCPPs in both UPF file format (used by QUANTUM ESPRESSO) and psp8 file format (used by ABINIT). 

We then
create two {\tt Atom} objects {\tt a1} and {\tt a2} representing the Si and H atoms. These objects are then placed in an {\tt atomlist} array using one of MATLAB's array creation syntax.  For systems containing a large number of atoms, we can also use MATLAB's loop scripting capability to build such an array using, e.g.,
\begin{lstlisting}[language=Matlab, caption=Building an atom list array~\label{code:atomlist}]
a1 = Atom('Si');
a2 = Atom('H');
atomlist(1) = a1;
for j = 2:5
  atomlist(j) = a2;
end;
\end{lstlisting} 

We then specify the Cartesian coordinates for each atom as a $5 \times 3$ array {\tt xyzlist}.  In this example, these atomic coordinates are calculated from the reduced coordinates specified in ({\tt redxyz}) and the supercell defined by the matrix $C$. But it is possible to specify these coordinates directly.

In order to solve the Kohn-Sham problem associated with this molecule,  we must also specify the kinetic energy cut-off {\tt ecut} to be used for the plane-wave discretization of the Kohn-Sham orbitals.  In the Listing~\ref{code:setup}, {\tt ecut} is set to 12.5 Hartree. 

All attributes of the $\mathrm{SiH}_4$ molecule are passed into the function that creates a {\tt Molecule} object as key--value pairs as shown in Listing~\ref{code:setup}.

Once a molecule object has been properly defined, we can solve the Kohn-Sham problem associated with this molecule by calling the {\tt scf} function as
\begin{verbatim}
[mol,H,X,info] = scf(mol);
\end{verbatim}

Running the {\tt scf} function generates the output shown in Listing~\ref{code:SCF output}.
\begin{lstlisting}[language=Matlab, caption=SCF Output~\label{code:SCF output}]
Beging SCF calculation for SiH4...
SCF iter   1:
eigtol =   1.000e-02
Rel Vtot Err    =            1.024e-01
Total Energy    = -6.2382906512612e+00
......
SCF iter  10:
eigtol =   9.543e-07
Rel Vtot Err    =            1.250e-06
Total Energy    = -6.2542498381078e+00
Elapsed time is 3.132360 seconds.
......
||HX-XD||_F     =            1.144e-08
\end{lstlisting} 

The default output written in the MATLAB command line window shows the convergence history of the SCF iteration, The output contains the dynamically adjusted error tolerance used to terminate iterative solution of a linear eigenvalue problem in each SCF iteration. It also contains the measurement of self-consistency error defined as
\begin{equation}
\frac{\| V_{\mathrm{in}} - V_{\mathrm{out}} \|}{\|V_{\mathrm{in}}\|},
\label{eq:scferr}
\end{equation}
where $V_{\mathrm{in}}$ is the sum of potential terms in \eqref{eq:ksham} that are functional of the electron density or density matrix at the beginning of each SCF iteration, and $V_{\mathrm{out}}$ is the corresponding new potential sum evaluated from the solution of the linear eigenvalue problem.  The Frobenius norm of the eigenpair residual $ H X - X \Lambda$ is also printed out, where $X$ contains all the desired eigenvectors and $\Lambda$ is a diagonal matrix containing the corresponding eigenvalues.

\subsection{Visualization and Post-processing}\label{sec:visuse}
In addition to the interactive output displayed in MATLAB's command line window, the {\tt scf} function also returns a number of output variables that can be further examined and visualized.  The returned {\tt Molecule} object (which in the example given here overwrites the input argument {\tt mol} includes the atomic forces computed for each atom at the end of the SCF iteration. We can examine these forces simply by typing
\begin{verbatim}
   mol.xyzforce
\end{verbatim}
on the command line, which produces
\begin{verbatim}
ans =

    0.0000    0.0000    0.0000
    0.0021    0.0021    0.0021
   -0.0021   -0.0021    0.0021
    0.0021   -0.0021   -0.0021
   -0.0021    0.0021   -0.0021
\end{verbatim}

\begin{figure}[ht]
\begin{center}
\includegraphics[width=0.5\textwidth]{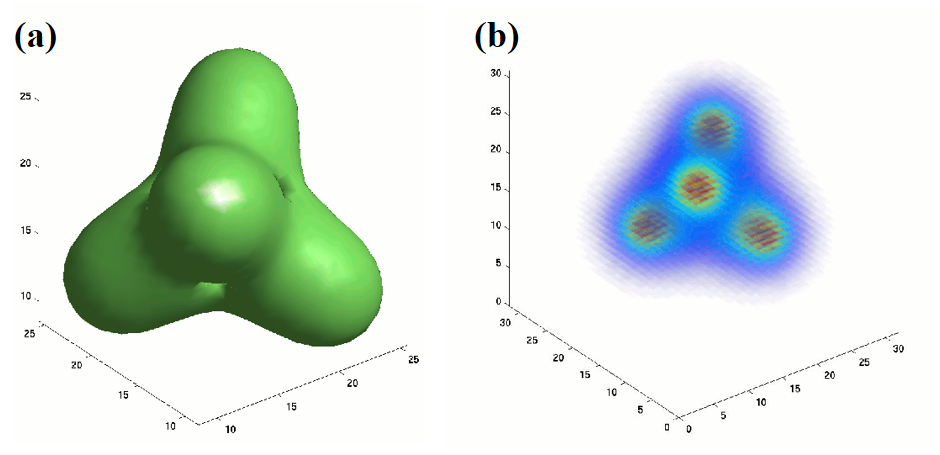}
\end{center}
\caption{(a) An isosurface rendering of the converged electron density of $\mathrm{SiH}_4$.
(b) A volume rendering of the converged electron density of $\mathrm{SiH}_4$.}
\label{fig:showrho}
\end{figure}

The returned Hamiltonian object {\tt H} contains the electron density $\rho$ as one of its attributes, 
which we can visualize by using a third party volume rendering function {\tt vol3d} included in KSSOLV or simply MATLAB's isosurface rendering function {\tt isosurface} as shown in Listing~\ref{code:showrho}. The {\tt fftshift} function used in the listing is called to re-center $\rho$ to the middle of the unit cell (instead of the origin of the Cartesian grid).
\begin{lstlisting}[language=Matlab, caption=Visualize the electron density~\label{code:showrho}]
view(3);
isosurface(fftshift(H.rho));
figure;
view(3);
vol3d('cdata',fftshift(H.rho));
\end{lstlisting} 
These renderings are shown in Figure~\ref{fig:showrho}.

We can also show a squared amplitude of Kohn-Sham orbital $\psi_j$. This
requires some post-processing of the returned {\tt Wavefun} object ${\tt X}$.
The post-processing involves using FFT to transform the default compact reciprocal space representation of the wavefunction {\tt XG} to a real space vector representation {\tt XR}, evaluating its magnitude square as {\tt $\mathrm{abs(XR).}^2$}, and reshaping the resulting vector into a 3D array. KSSOLV provides a utility function {\tt poscar} to write the magnitude square of the reshaped wavefunction to a text file that can be read by other visualization software tools such as the VESTA~\cite{vesta}.

\begin{figure}[ht]
\begin{center}
\includegraphics[width=0.5\textwidth]{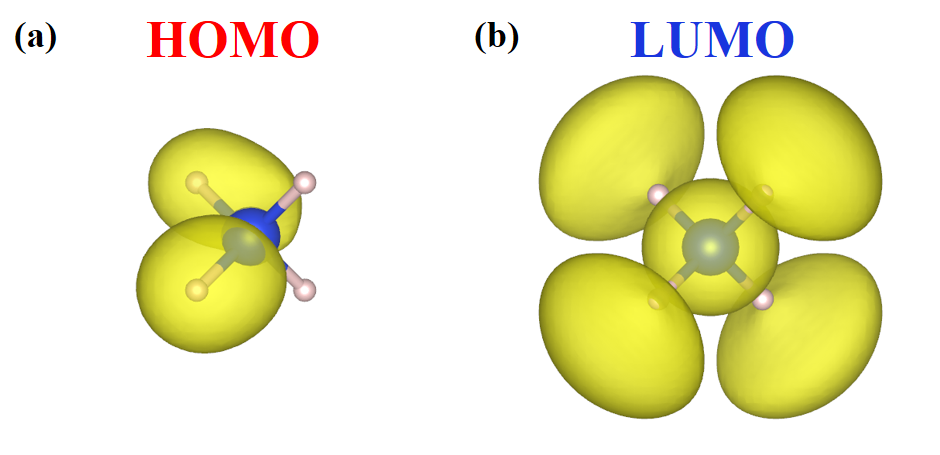}
\end{center}
\caption{The isosurfaces of HOMO and LUMO produced by VESTA. (a) HOMO of $\mathrm{SiH}_4$.
(b) LUMO of $\mathrm{SiH}_4$.}
\label{fig:showhomolumo}
\end{figure}

Listing~\ref{code:homo} shows how post-processing is performed to write the magnitude square of the highest occupied molecular orbital (HOMO) to a file named {\tt SiH4\_HOMO}. A similar set of commands can be used to write the lowest unoccupied molecular orbital (LUMO) to another file. The HOMO and LUMO can be subsequently visualized by using the VESTA software as shown in Figure~\ref{fig:showhomolumo}.

\begin{lstlisting}[language=Matlab, caption= Wavefunction post-processing and visualization of the HOMO~\label{code:homo}]
n1=mol.n1;n2=mol.n2;n3=mol.n3;
homo = mol.nel/2;
lumo = homo + 1;
XG = X.psi(:,homo);
F = KSFFT(mol);
XR = (F'*XG)*sqrt(mol.vol);
X2 = abs(XR).^2;
pos = poscar(mol);
outchg('SiH4_HOMO', pos, reshape(X2,n1,n2,n3));
\end{lstlisting}

The returned {\tt info} argument is a MATLAB structure that contains several fields.
\begin{verbatim}
>> info

info = 

  struct with fields:

      Eigvals: [4×1 double]
      Etotvec: [10×1 double]
    SCFerrvec: [10×1 double]
         Etot: -6.2542
\end{verbatim}

We can plot the convergence history of the SCF iteration by simply using 

\begin{verbatim}
x = [1:length(options.maxscfiter)]
fig1 = semilogy(info.SCFerrvec_lob,'-s'
       ,info.SCFerrvec_dia,'-d');
legend([fig1(1)fig1(2)],{'LOBPCG+Anderson'
       ,'Davidson+Broyden'});
xlabel('SCF iteration number','FontName',
       'Times New Roman') 
ylabel('SCF error','FontName',
       'Times New Roman')
set(gca,'XTick',x)
\end{verbatim}

\begin{figure}[ht]
\begin{center}
\includegraphics[width=0.48\textwidth]{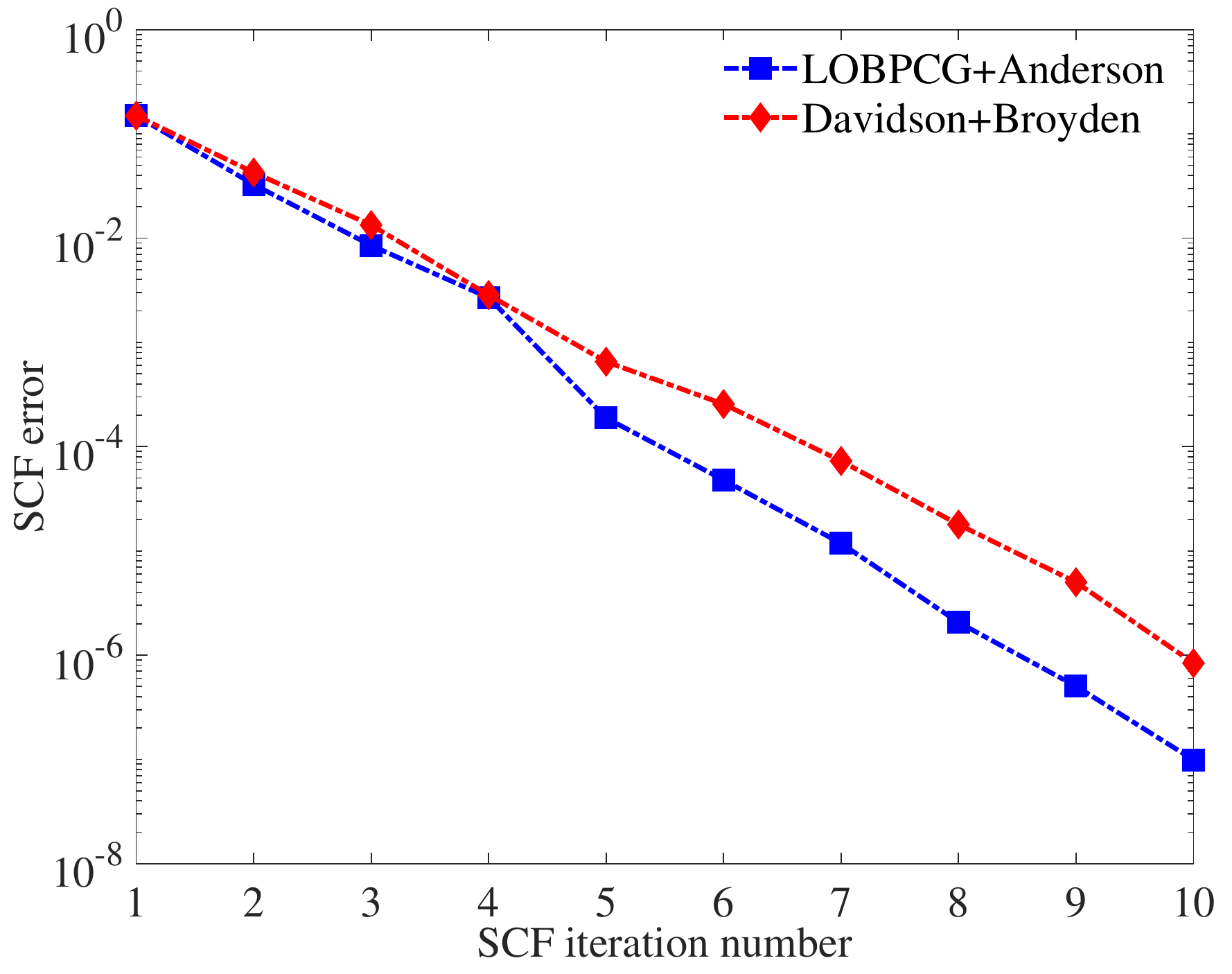}
\end{center}
\caption{The change of SCF error \eqref{eq:scferr} with respect to SCF iteration number. Two combinations of diagonalization algorithm and mixing method are given, 1. LOBPCG with Anderson, 2. Davidson with Broyden.}
\label{fig:scferr_10}
\end{figure}

\begin{figure}[ht]
\begin{center}
\includegraphics[width=0.5\textwidth]{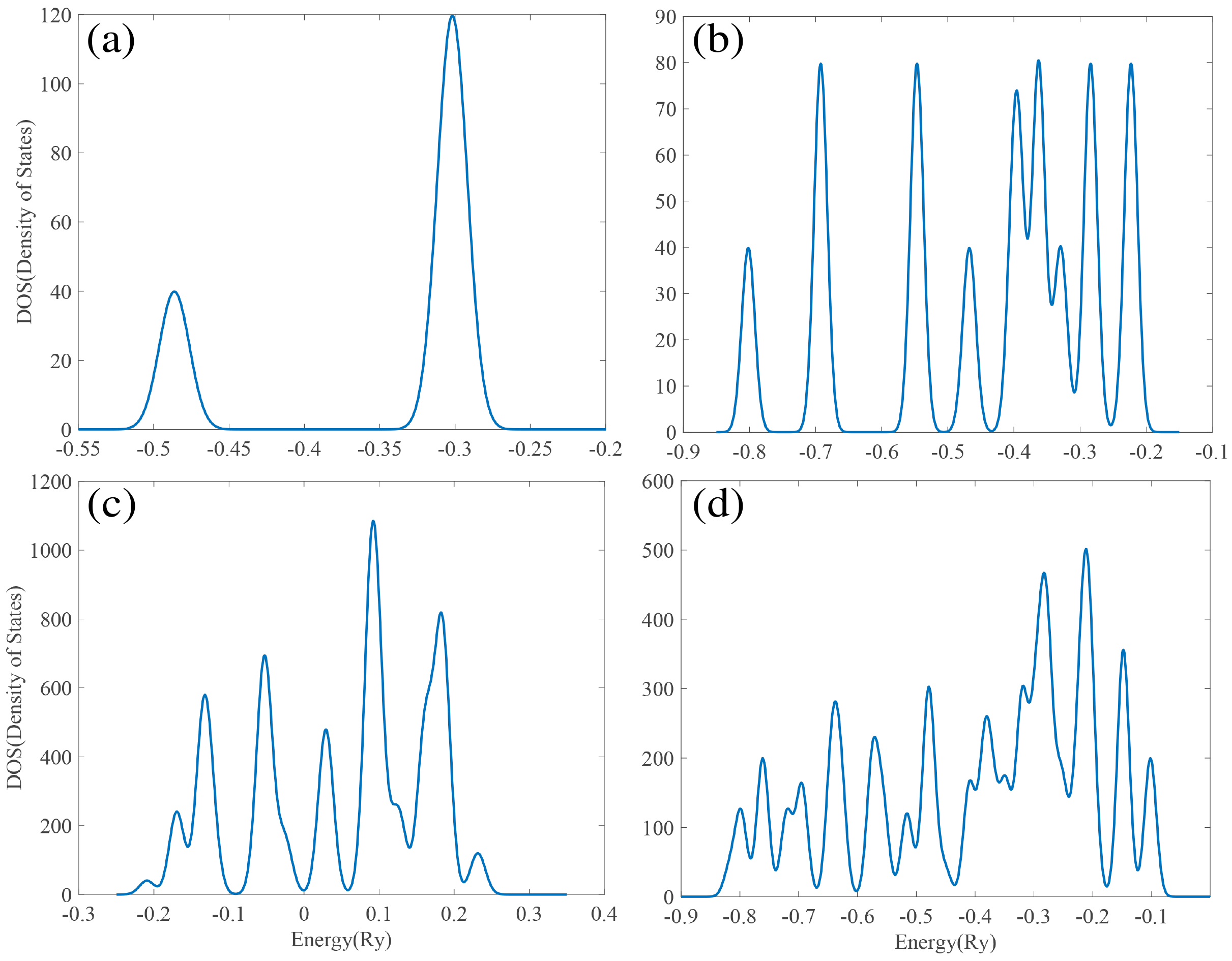}
\end{center}
\caption{DOS(Density of States) by using Gaussian function, which describes the proportion of states that are to be occupied by the system at each energy. (a)SiH$_{4}$, (b)C$_{6}$H$_{6}$, (c)Si$_{64}$, (d)C$_{60}$. }
\label{fig:dos}
\end{figure}

We can also use the {\tt Eigvals} information from the {\tt info} argument to obtain the DOS(Density of States). The post-processing includes setting some parameters to get the energy range and using either the Gaussian or the Lorentzian spread function to create a smooth DOS curve from {\tt info.Eigvals}.
Listing~\ref{code:dos} gives a simple script for carrying out such type of post-processing. The DOS curves produced for four different systems (SiH$_{4}$, C$_{6}$H$_{6}$, Si$_{64}$ and C$_{60}$) are shown in Figure~\ref{fig:dos}.
\begin{lstlisting}[language=Matlab, caption=Energy post-processing and visualization of the DOS(density of states.)~\label{code:dos}]
ev = info.Eigvals;
nx = 1000;
sigma = 0.01;
% Initialization 
[m,n]   = size(ev);
ne      = m;
emin    = -1;%min(ev);
emax    = 1;%max(ev);
xgrid   = (emax - emin)/(nx - 1);
dos     = zeros(nx,2);
% Get the x distribution of the energy
for ix = 1 : nx
  dos(ix,1) = emin + (ix - 1)*xgrid;
end
% Calculate DOS
for ie = 1 : ne
  for ix = 1 : nx
    x = emin + (ix - 1)*xgrid - ev(ie);
    %if(Gaussian)
    dos(ix,2) = dos(ix,2) + 1/(sigma*sqrt(2*pi))
    *exp(-x^2/(2*sigma^2));
    %if(Lorentzian) 
    %dos(ix,2) = dos(ix,2) + sigma/(pi*(x^2+sigma^2));
  end
end
\end{lstlisting} 

\subsection{Modifying options and algorithms}
KSSOLV 2.0 allows users to choose and experiment with different algorithms or algorithmic components for solving the Kohn-Sham problem.  For example, instead of calling the {\tt scf} function, one can call the {\tt trdcm} function, which implements the trust region regularized DCM algorithm discussed in section~\ref{sec:scfdcm}, as
\begin{verbatim}
[mol,H,X,info] = trdcm(mol);
\end{verbatim}

Both the {\tt scf} and {\tt trdcm} functions accept an additional option argument that allows users to alter the default algorithms and parameters used by these functions.

The optional argument can be first created by calling the {\tt setksopt} function which returns a MATLAB structure that contains default algorithmic choices and parameters listed in Listing~\ref{code:setksopt}.
\begin{lstlisting}[language=Matlab, caption=option structure returned from the {\tt setksopt} function~\label{code:setksopt}]
           verbose: 'off'
         eigmethod: 'lobpcg'
        maxscfiter: 10
        maxdcmiter: 10
        maxinerscf: 3
         maxcgiter: 10
       maxeigsiter: 300
            scftol: 1.0000e-08
            dcmtol: 1.0000e-08
             cgtol: 1.0000e-09
           eigstol: 1.0000e-10
          what2mix: 'pot'
           mixtype: 'anderson'
            mixdim: 9
           betamix: 0.8000
             brank: 1
                X0: []
              rho0: []
            degree: 10
             force: 1
          ishybrid: 0
            useace: 0
              Vexx: []
        maxphiiter: 5
            phitol: 1.0000e-08
            dfrank: 0
           ncbands: 0
       relaxmethod: 'fminunc'
          relaxtol: 1.0000e-04
    factorOrbitals: 1
          davsteps: 3
           ngbands: 0
\end{lstlisting} 

We can change, for example, the algorithm for solving the linear eigenvalue problem in each SCF iteration from LOBPCG to Davidson, and the quasi-Newton algorithm (charge mixing scheme) used to accelerate the SCF iterations from Anderson to Broyden by using the commands given in Listing~\ref{code:algo} to
modify the {\tt options} structure and passing it to the {\tt scf} function along with the {\tt mol} object.
\begin{lstlisting}[language=Matlab, caption=Choosing a different eigensolver and charge mixing scheme.~\label{code:algo}]
options = setksopt();
options.eigmethod = 'davidson';
options.mixtype = 'broyden';
[mol1,H1,X1,info1] = scf(mol,options);
\end{lstlisting} 

Figure~\ref{fig:scferr_10} shows these changes lead to a slightly difference convergence behavior of the SCF iteration (the red curve) although the difference is relatively small in this particular case.

We can see from Figure~\ref{fig:scferr_10} that the SCF iteration did not converge to the default accuracy requirement specified by the parameter {\tt options.scftol}, which is set to $10^{-8}$. To reach that level of accuracy, we can rerun the {\tt scf} function by using the wavefunction {\tt X} and electron density {\tt rho} returned from the previous run as the starting guess. This can be achieved by simply setting {\tt options.X0} and {\tt options.rho0} to the previously returned wavefunction and electron density.
\begin{verbatim}
    options.X0 = X;
    options.rho0 = H.rho;
\end{verbatim}

After calling the {\tt scf} function with the modified option as an input, we can reach convergence as reported in Listing~\ref{code:sih4conv}.

\begin{lstlisting}[language=Matlab, caption=Convergence is reached after rerunning {\tt scf} with the wavefunction and electron density initialized to the approximation produced from the first {\tt scf} call.~\label{code:sih4conv}]
Regular SCF for Pure DFT
Beging SCF calculation for SiH4...
SCF iter   1:
eigtol =   1.000e-02
Rel Vtot Err    =            1.146e-07
Total Energy    = -6.2542498381078e+00
...
SCF iter   4:
eigtol =   1.257e-09
Rel Vtot Err    =            4.190e-09
Total Energy    = -6.2542498381079e+00
Convergence is reached!
Elapsed time is 1.005311 seconds.
Etot            = -6.2542498381079e+00
Eone-electron   = -5.3304963587462e+00
Ehartree        =  3.2198596360275e+00
Exc             = -2.5983987591201e+00
Eewald          = -1.5452143562691e+00
Ealphat         =  0.0000000000000e+00
--------------------------------------
Total time used =            4.563e+00
||HX-XD||_F     =            2.562e-09
\end{lstlisting} 

\subsection{Algorithm prototype and modification}\label{sec:motiv}

KSSOLV is designed to enable researchers to easily modify existing algorithms 
and prototype new algorithms. To a large extent, this feature is facilitated 
by the object-oriented programming model supported in MATLAB. By creating Hamiltonian and wavefunction objects and overloading the basic linear algebra 
operations with these objects as operands, one can literally translate
mathematical expressions into MATLAB codes in KSSOLV in a few minutes.
To give an example, let us take a look at the implementation
of the ACE operator for hybrid functional DFT calculation in KSSOLV which is 
shown in Listing~\ref{code:ace}.
\begin{lstlisting}[language=Matlab, caption=Constructing the ACE operator~\label{code:ace}]
    W = ApplyVexx(X);
    M = X' * W;
    M = (M + M')/2;
    R = chol(-M);
    Y = W / R;
    ApplyVexxACE = @(x) -Y * (Y' * x);
\end{lstlisting} 

The ACE operator is defined by \eqref{eq:Vace} which, once $\psi_j(\br)$'s are discretized and
represented by columns of the matrix $X$, can also be written in matrix form as
\begin{equation}
\hat{V}^{\mathrm{ACE}} = -W M^{-1} W^{\ast},
\label{eq:vacemat}
\end{equation}
where
\begin{equation}
W = \hat{V}^{\mathrm{HSE}}(X) X,
\label{eq:applyvhse}
\end{equation}
with $\hat{V}^{\mathrm{HSE}}(X)$ being the matrix representation of the Hartree-Fock exchange operator 
and $M = X^{\ast} W$. Because $-M$ is Hermitian positive definite, we can rewrite \eqref{eq:vacemat} in 
a symmetric form by performing a Cholesky factorization of $-M$, i.e., $-M = RR^{\ast}$ with $R$ being 
upper triangular, and expressing $\hat{V}^{\mathrm{ACE}}$ as $\hat{V}^{\mathrm{ACE}} = -YY^{\ast}$ with 
$Y = WR^{-1}$.

In Listing~\ref{code:ace}, we apply the function {\tt ApplyVexx}, which 
implements \eqref{eq:applyvhse}, to the {\tt Wavefun} object {\tt X} to obtain another {\tt Wavefun}
object {\tt W}. Even though {\tt X} and {\tt W} are {\tt Wavefun} objects, we can treat
them as matrices and multiply them together in the second line of Listing~\ref{code:ace} to obtain
the matrix $M=X^{\ast} W$. Line 3 in Listing~\ref{code:ace} is used to ensure {\tt M} is numerically Hermitian 
before the Cholesky factorization function {\tt chol} is applied to {\tt -M}. The inverse of the Cholesky
factor {\tt R} is applied to {\tt W} to yield the {\tt Wavefun} object {\tt Y} by solving a set of linear
equations using the MATLAB {\tt /} operator. The {\tt Y} object is then used to define a function handle 
{\tt ApplyVexxACE} that can be applied to any {\tt Wavefun} object of matching dimensions without explicitly 
forming the ACE operator.

\subsection{Performance profiling}
MATLAB provides a convenient performance profiling tool that allows us to easily analyze the performance features of KSSOLV functions and identify potential computational bottlenecks.  
For example, to profile the performance of the HSE06 calculation contained
in a testing script named {\tt testHSE06.m}, we can simply issue the
following several commands listed in Listing~\ref{code:testHSE06}.
\begin{lstlisting}[language=Matlab, caption=Profiling for a HSE06 calculation.
%\CY{Can we replace this with HSE06 profiling?}
~\label{code:testHSE06}]
profile on; 
testHSE06;
save profHSE06 p
profile off;
\end{lstlisting} 

MATLAB provides a viewer in the Windows Visual interface that allows us to clearly see the hierarchical relationship among different computational components as well as which function takes most of the time. We can further zoom into the most time-consuming function and
identify the line number of the code that takes most of the time within that function.

\begin{figure*}[ht]
\begin{center}
\includegraphics[width=1\textwidth]{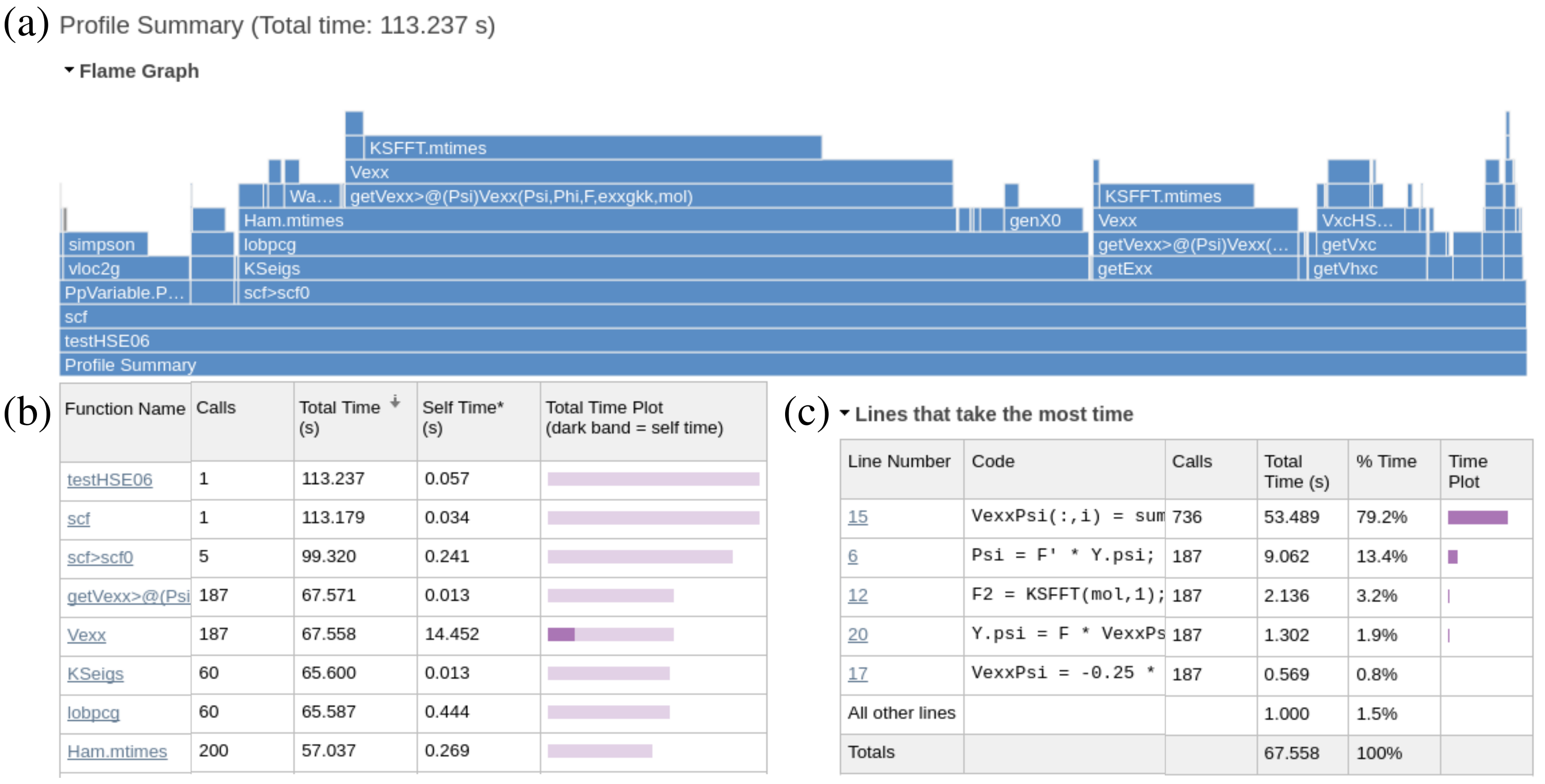}
\end{center}
\caption{Profiling of a HSE06 calculation, (a)The overall profile summary flame graph, (b)Functions hot spot analysis, including the called times of sub-functions, (c)Time corresponding to each line of {\tt getVexx}.
}
\label{fig:HSEprof}
\end{figure*}

For example,  Figure~\ref{fig:HSEprof}(a)  displays a Flame graph of KSSOLV functions called by {\tt testHSE06}, the most time consuming function is {\tt getVexx}, which is used to calculate the exchange potential. By clicking on the block containing this function name, we obtain a table shown in Figure~\ref{fig:HSEprof}(b), which lists the line numbers of the most time consuming functions contained in {\tt getVexx}. If we click the line number associated with a particular function, we can step in the code of that function and analyze the computation performed in that function. 

Once we have identified the computational bottleneck of HSE06 calculation, which is in the evaluation of the exchange term, we can optimize the performance of KSSOLV by seeking alternative implementations of the functions in question or using alternative algorithms. For example, as we discussed in section~\ref{sec:C} and \ref{sec:motiv}, the use of the ACE algorithm to refactor the exchange operator can significantly 
reduce the computational complexity of applying the Hartree-Fock exchange 
operator to a set of wavefunctions in the hybrid functional DFT calculation.
Table~\ref{tab:tabace} gives a direct comparison between the cost of hybrid functional DFT calculations with and without the use of ACE. We can clearly see that after using ACE, the total amount of wall clock time used by {\tt scf0} is reduced from 76 seconds to 28 seconds. This is mainly due to a significant reduction in time spent in the {\tt lobpcg} eigensolver used for the outer iteration.  With the use of ACE, the function {\tt getVexx} which is used to update the Fock exchange operator, is called only 5 times (in the 5 outer {\tt scf0} iterations), whereas 187 such calls are made in the hybrid functional DFT calculation without using ACE. Furthermore, the use of the ACE allows us to significantly reduce the number of FFTs used to apply the Fock exchange operator to a set of wavefunctions.  In the ACE enabled hybrid functional calculation, only 55 times FFTs are used to construct the ACE operator, whereas 1851 FFTs are performed when the Fock exchange operator is applied to a set of wavefunctions in each step of the LOBPCG eigensolver. Table~\ref{tab:tabace} also shows that the overhead incurred in constructing the ACE operator is relatively small, i.e. 1.5 seconds (used by {\tt calculateACE}) out of 28 seconds used by {\tt scf0}.

\begin{table}[!tb]\footnotesize
\centering
\setlength{\tabcolsep}{3mm}
\caption{Cost comparison between HSE and HSE-ACE calculations. The system calculated here is SiH$_{4}$ with $E_{\rm cut}$ set to 20 Hartree.} 
\begin{tabular}{ccc} \ \\
\hline \hline
Function name &  Calls numbers & Time(s) \ \\
\hline
{\tt scf0}(HSE) & 5 & 75.965 \ \\
{\tt lobpcg}(HSE) & 59 & 48.779 \ \\
{\tt getVexx}(HSE) & 187 & 49.969 \ \\
{\tt KSFFT.mtimes}(HSE) & 1851 & 38.448 \ \\
\hline
{\tt scf0}(HSE-ACE) & 5 & 28.055 \ \\
{\tt lobpcg}(HSE-ACE) & 59 & 13.778 \ \\
{\tt getVexx}(HSE-ACE) & 5 & 1.554 \ \\
{\tt KSFFT.mtimes}(HSE-ACE) & 55 & 1.362 \ \\
{\tt calculateACE}(HSE-ACE) & 5 & 1.554 \ \\  
\hline \hline
\end{tabular}
\label{tab:tabace}
\end{table}

\section{Results and discussion}\label{sec:results}
In this section, we give some examples of a few applications that can be studied with KSSOLV and demonstrate its accuracy and performance. The descriptions of these systems are listed in Table~\ref{tab:benchset}.

\subsection{Accuracy}
We first use KSSOLV to perform ground state total energy and atomic force calculations, band structure analysis and geometry optimization for a few molecules and solids.
In all these runs,  we set the inner SCF convergence tolerance to $10^{-7}$ for calculations that use LDA and PBE functionals, and $10^{-6}$ for outer SCF convergence tolerance when using the HSE06 functional. We use QUANTUM ESPRESSO as the baseline for comparison in assessing the accuracy of KSSOLV.

\begin{table*}[!tb]\footnotesize
\centering
\setlength{\tabcolsep}{1mm}
\caption{The performance of KSSOLV on a set of test problems. System: including solid, molecule, nanotube, $n_a$: number of atoms, Cell dim: the unit cell size with three dimensions, $N_{e}$: number of electrons, $n_k$: number of k-points, Functional: exchange-correlation functional, Ecut: cut-off energy, $N_{r}$: grids number in real space, $N_{g}$: grids number in reciprocal space, Scftol : the converge limit of inner SCF, Scf-iter: the iteration number of inner SCF, Phitol : the converge limit of outer SCF, Phi-iter: the iteration number of outer SCF, Total time: the total wall clock time of each calculation.}
\begin{tabular}{cccccccccccccc} \ \\
\hline \hline
System &  $n_a$  & Cell dim  & $N_{e}$ & $n_k$ & Functional & Ecut (Ha) & $N_{r}$ & $N_{g}$ & Scftol & Scf-iter & Phitol & Phi-iter & Total time (sec)\ \\
\hline
\textbf{SiH$_{4}$}   &  5 & $20$ & 8 & 1 & PBE & 20 & 531441 & 34265 & $10^{-7}$ & 14 & - & - & 17.544 \ \\
\textbf{C$_{6}$H$_{6}$}   &  12 & $22.4\times 24.9 \times 30.2$ & 30 & 1 & PBE & 20 & 1121302 & 72079 & $10^{-7}$ & 18 & - & - & 154.358 \ \\
\textbf{Si$_{64}$}   &  64 & $20.52^3$ & 256 & 1 & PBE & 20 & 571,787 & 37,073 & $10^{-7}$ & 19 & - & - & 642.018 \ \\
\textbf{C$_{60}$}    &  60 & $24.57^3$ & 240 & 1 & PBE & 20 & 970,299 & 63,317 & $10^{-7}$ & 20 & - &  -  & 1174.029 \ \\
\textbf{Si$_{216}$}  & 216 & $30.78^3$ & 864 & 1 & PBE & 20 & 1906624  & 124289 & $10^{-7}$ & 19 & - &  -  & 8769.963\ \\
\textbf{Si$_{64}$}   &  64 & $20.52^3$ & 256 & 1 & HSE & 20 & 571,787 & 37,073 & $10^{-7}$ & 19 & $10^{-6}$ & 3 & 79077.64 \ \\
\textbf{Si$_{64}$}   &  64 & $20.52^3$ & 256 & 1 & HSE-ACE & 20 & 571,787 & 37,073 & $10^{-7}$ & 19 & $10^{-6}$ & 4 & 3493.511 \ \\
\textbf{C$_{60}$}   &  60 & $24.57^3$ & 240 & 1 & HSE & 20 & 970299 & 63317 & $10^{-7}$ & 20 & $10^{-6}$ & 3 & 184672.234 \ \\
\textbf{C$_{60}$}   &  60 & $24.57^3$ & 240 & 1 & HSE-ACE & 20 & 970299 & 63317 & $10^{-7}$ & 20 & $10^{-6}$ & 4 & 5058.289 \ \\
\textbf{CNT661}   &  60 & $38\times 38\times 4.6$ & 96 & 1 & PBE & 20 & 399475 & 25485 & $10^{-7}$ & 19 & - & - & 146.720 \ \\
\textbf{Si$_{8}$}   &  8 & $10.216^3$ & 96 & 64 & PBE & 20 & 74088 & 4553 & $10^{-7}$ & 14 & - & - & 345.123 \ \\
\textbf{Cu$_{4}$}   &  4 & $6.8308^3$ & 32 & 64 & PBE & 30 & 39304 & 2517 & $10^{-5}$ & 18 & - & - & 512.685 \ \\
\hline \hline
\end{tabular}
\label{tab:benchset}
\end{table*}

\subsubsection{Total energy and atomic forces}\label{sec:Energy}

When comparing with the QE (QUANTUM ESPRESSO) results, we compute the total energy difference per atom as well as the maximum difference in atomic forces, which are defined by
\begin{eqnarray*}
\Delta E&=&\left(E^{\mathrm{KSSOLV}}_\mathrm{tot}-E^{\mathrm{QE}}_\mathrm{tot}\right) / N_{\mathrm{A}}, \\
\Delta F&=&\max _{I}\left\|F_{I}^{\mathrm{KSSOLV}}-F_{I}^{\mathrm{QE}}\right\|,
\end{eqnarray*}
where $E^{\mathrm{KSSOLV}}_\mathrm{tot}$ and $E^{\mathrm{QE}}_\mathrm{tot}$ are converged total energy levels returned from KSSOLV and QE respectively, $N_{A}$ is the total number of atoms in each system,  and $I$ is an atom index.

\begin{figure}[ht]
\begin{center}
\includegraphics[width=0.5\textwidth]{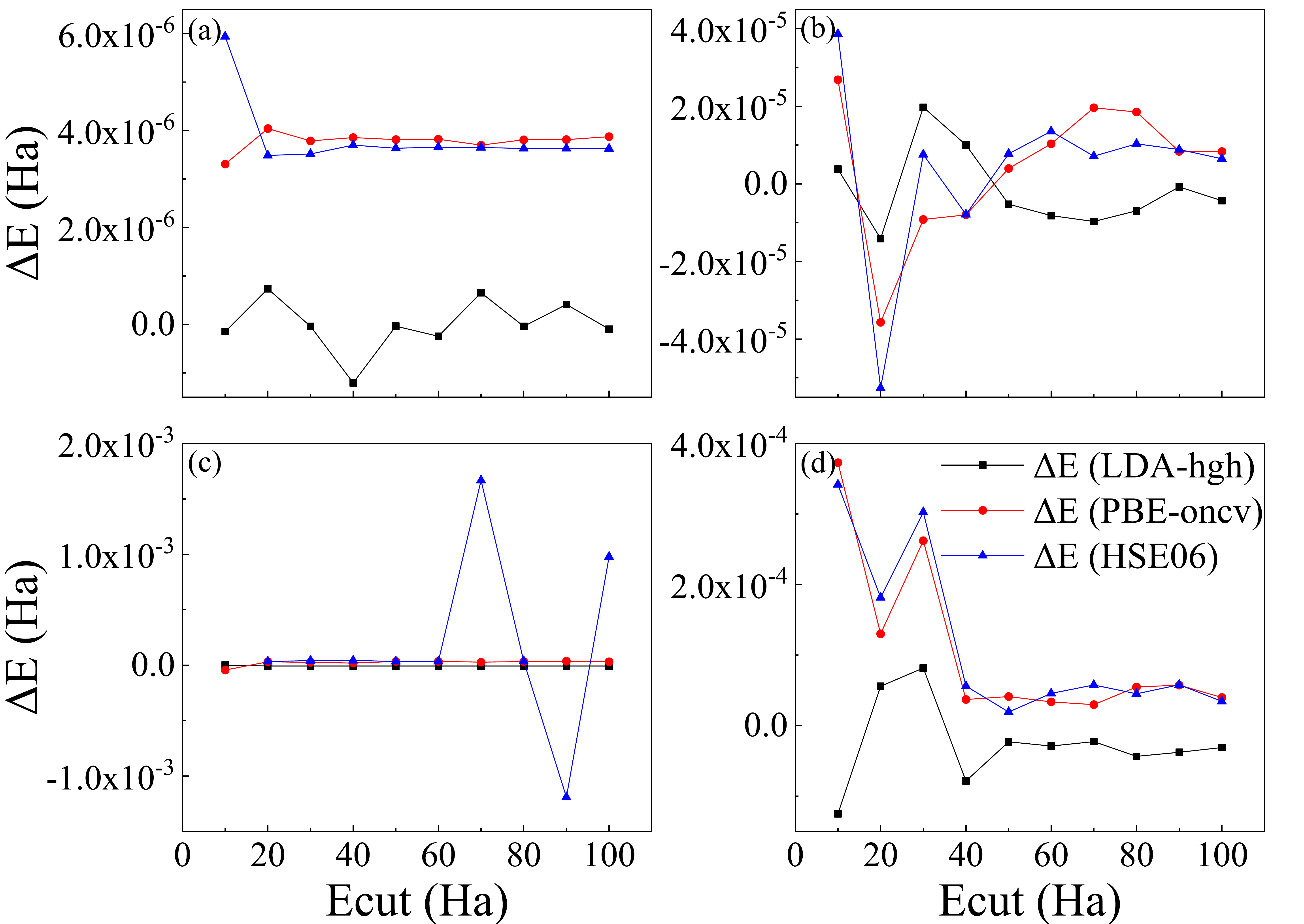}
\end{center}
\caption{Total energy difference between KSSOLV and QE for (a) silane (SiH$_{4}$) molecule, (b) benzene (C$_{6}$H$_{6}$) molecule, (c) bulk silicon Si$_{64}$ and (d) fullerene (C$_{60}$) molecule at different plane-wave cut-off energy levels.}
\label{fig:energy}
\end{figure}

To check the accuracy systematically, we measure $\Delta E$ and $\Delta F$ for each system at several plane-wave cut-off energy ($E_{cut}$) levels (from 10 to 100 Hartree). We also
use three types of psedopotential and exchange-correlation functional combinations in \kssolv{} 2.0, which are LDA-HGH, PBE-ONCV and HSE06, respectively.
The total energy differences for test systems are plotted in Figure~\ref{fig:energy}. 
The solid black square lines correspond to the LDA-HGH exchange-correlation functional and psedopotential combination, the solid red circle lines correspond to PBE-ONCV, and the solid blue triangle lines correspond to HSE06. 

We observe that, in general, the difference between the converged KSSOLV and QE total energies per atom is on the order of between $10^{-6}$ and $10^{-4}$ Hartree, which is sufficiently small. For Si$_{64}$, the energy difference is slightly larger (on the order of $10^{-3}$ Hartree) at some plane-wave cut-off levels. However, these differences are acceptable since they are around chemical accuracy, which is defined to be 1kcal/mol or 10$^{-3}$ Hartree, and are sufficient for most applications.
In previous studies~\cite{gace}, we also compared differences in cohesive energies for several test problems and showed that they match well.

\begin{figure}[ht]
\begin{center}
\includegraphics[width=0.5\textwidth]{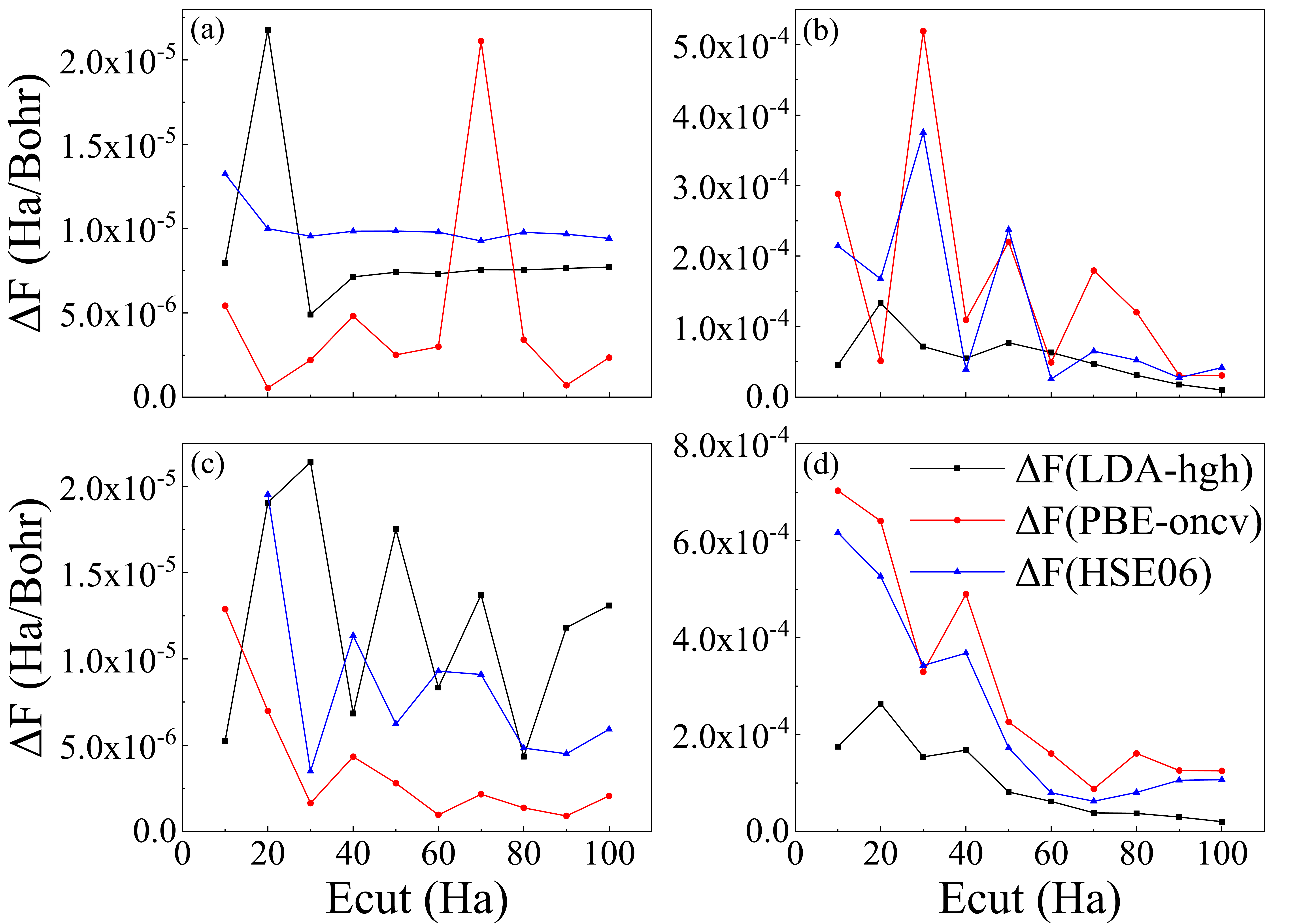}
\end{center}
\caption{Maximum difference in atomic forces between KSSOLV and QE for (a) silane (SiH$_{4}$) molecule, (b) benzene (C$_{6}$H$_{6}$) molecule, (c) bulk silicon Si$_{64}$ and (d) fullerene (C$_{60}$) molecule at different plane-wave cut-off energy levels.} \label{fig:atomforce}
\end{figure}

In Figure \ref{fig:atomforce}, we plot the maximum difference in atomic forces between KSSOLV and QE for all test systems. The magnitude of force difference is generally small and within the range of $10^{-6}$ to $10^{-4}$ Hartree/Bohr. In some cases, the difference is slightly larger when $E_{\mathrm{cut}}$ is relatively small, but becomes sufficiently small when $E_{\mathrm{cut}}$ reaches 50 Ha or so. 

\subsubsection{Band Structure}
\begin{figure}[ht]
\begin{center}
\includegraphics[width=0.5\textwidth]{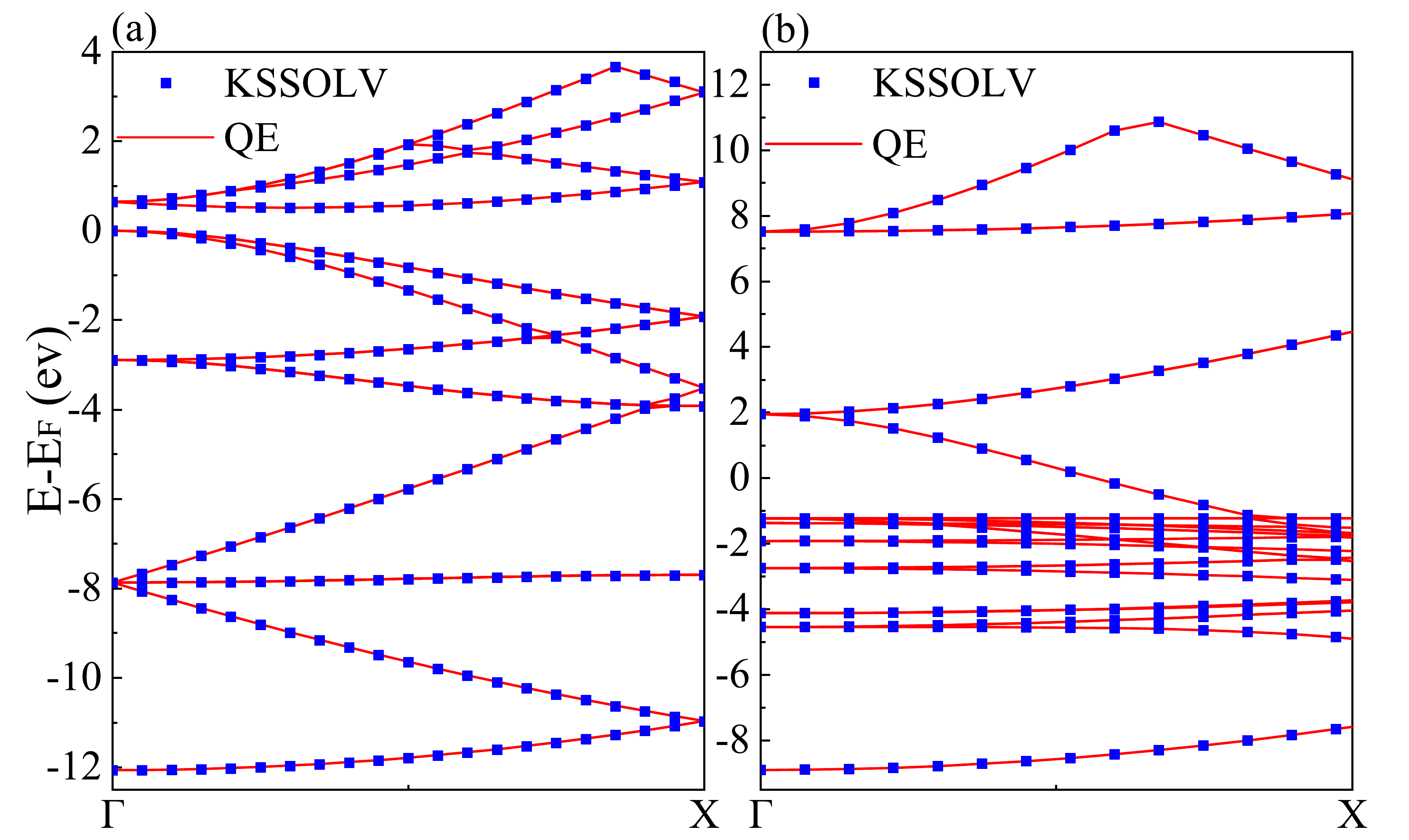}
\end{center}
\caption{The band structures calculated by KSSOLV and QE using the PBE functional for (a) bulk silicon Si$_{8}$ and (b) bulk copper Cu$_{4}$ system. The cut-off energies are 20 Hartree and 30 Hartree for Si$_{8}$ and  Cu$_{4}$, respectively. All energy levels have been shifted to keep the Fermi energy at zero ev.} \label{fig:band}
\end{figure}

Because KSSOLV 2.0 facilitates $k$ point samplings in the first Brillouin zone for solids, we 
can use it to compute band structures of solids and compared them with the results obtained from QE also.  In Figure~\ref{fig:band}, we plot the band structure of Si$_{8}$ and Cu$_{4}$ respectively between the $\Gamma$ and $\mathrm{X}$ $k$-points. In both cases,  
the number of k-points used in the SCF calculation is 64(4x4x4), 
we observe that the band structures obtained from KSSOLV are in excellent agreement with those obtained from QE, and the average numerical difference of band structure between the two packages is about $10^{-9}$. We can clearly see a band gap between the the highest valence
band and the lowest conducting band for Si$_{8}$ which confirms the fact that Si is a semiconductor.  No band gap can be seen in Figure~\ref{fig:band}(b) for Cu$_{4}$, this is consistent with the previous knowledge that Cu is a metal.

\subsubsection{Geometry optimization}
\begin{figure}[ht]
\begin{center}
\includegraphics[width=0.5\textwidth]{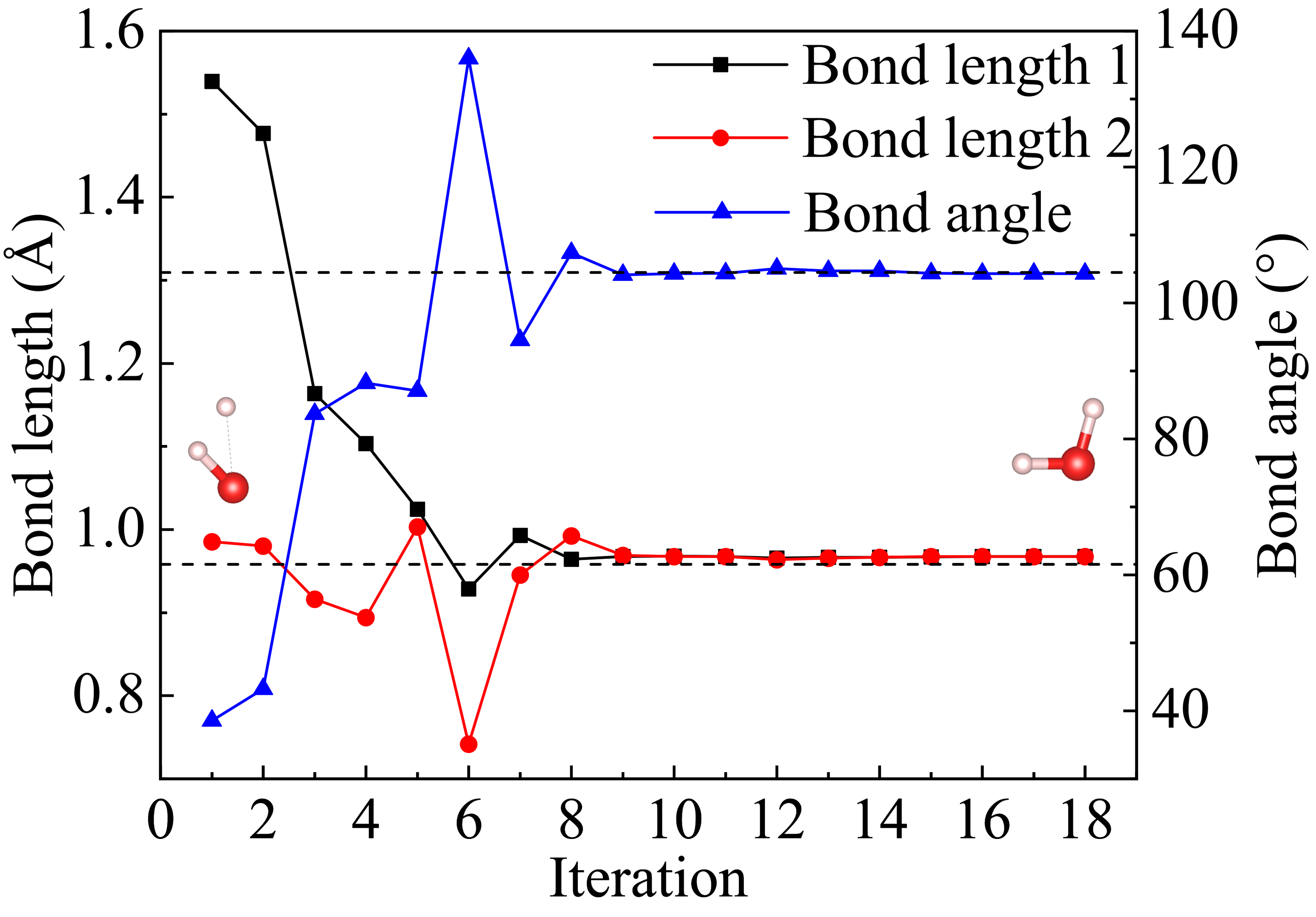}
\end{center}
\caption{Geometric optimization for H$_{2}$O, the solid black square line, the solid red dotted line, and the blue triangle represent two bond lengths and bond angle respectively, the black dotted line represents the experimental value. The total number of iteration steps is 18 and the initial and final structures are given at the first and the last iteration step. The cut-off energy of the calculation is 60 Hartree.} \label{fig:h2oopt}
\end{figure}

We use KSSOLV 2.0 to optimize the geometry of an isolated water (H$_2$O) molecule and compared the optimal H-O bond length, and the optimal angle between two H-O bonds with the corresponding experiment values. The initial bond lengths between the H and O atoms are set to 0.98523 and 1.53953 \AA, respectively, and the initial bond angle is set to 38.5624$^{\circ}$. These are slightly different from the experiment value~\cite{milovanovic2020flexible} of 0.957 $\AA$ for the bond length and 104.5$^{\circ}$ for the bond angle.  The BFGS algorithm implemented in MATLAB Optimization toolbox function {\tt fminunc} is used to perform the optimization. The convergence of bond length and bond angle to the experimentally observed values in the KSSOLV geometry optimization function {\tt relaxatoms} is shown in Figure~\ref{fig:h2oopt}. We can clearly see that convergence is reached after 9 geometry optimization steps. And the error of the  KSSOLV calculation result differs from the experimental value by only one percent.

\begin{figure}[ht]
\begin{center}
\includegraphics[width=0.5\textwidth]{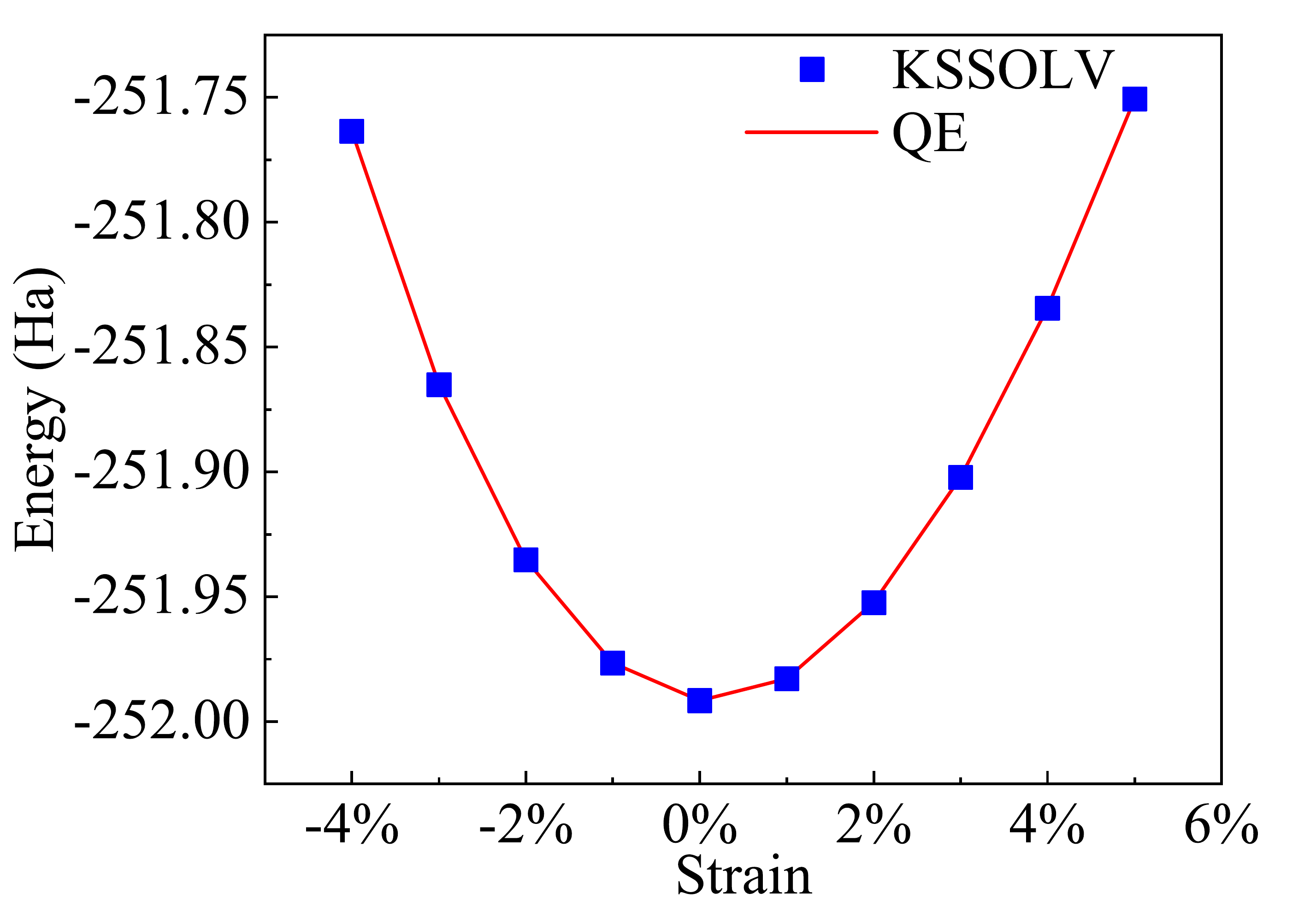}
\end{center}
\caption{Geometric optimization for bulk silicon Si$_{64}$ system, the energy change with strain, KSSOLV (the blue square), QE (solid red line), The unit of strain change is one percent of the overall structure, from compressive strain (-4$\%$) to tensile strain (+5$\%$), the cut-off energy of these calculations is 60 Hartree.} 
\label{fig:si64opt}
\end{figure}

To give another example, we perform a geometry optimization Si$_{64}$ with respect to the size of the unit cell.  The change in unit cell size corresponds to the change in the strain applied to the solid. To perform the optimization, we sample several unit cell sizes
that correspond to $0-6\%$ changes in applied strain, and compute the
ground state of Si$_{64}$ contained in these unit cells.  The results are
compared with those obtained from QE. Figure \ref{fig:si64opt} shows that
the KSSOLV results match well with QE results. These results clearly show the energy is the lowest at zero strain.

\subsection{Performance}
In this section, we report the performance of KSSOLV by applying it to a set of benchmark problems listed in Table~\ref{tab:benchset}.  The version of MATLAB we used is R2021a and the benchmark is run on a 
Intel(R) Xeon(R) CPU E5-2698 v4 @ 2.20GHz with 40 maximum threads. 
We perform ground state calculation for both molecules and solids using either PBE or HSE06 functionals 
specified in Table~\ref{tab:benchset}. For hybrid functional DFT calculations, the HSE exchange-correlation functional and a two-level SCF procedure are used. The standard SCF calculation is used as the inner iteration to reach self consistency in the electron density for a fixed Fock exchange potential. In the outer SCF iteration that we refer to as the Phi-iteration, the Fock exchange potential and energy are updated. Thus, two different convergence criteria are used in the inner and outer iteration.

We report the wall clock time it takes to complete the calculation as well as the number of SCF iterations required to reach convergence. The convergence criterion (i.e., the SCF error tolerance) for each case is listed in the table also.

The first four systems listed in Table~\ref{tab:benchset} were used in the previous subsections to demonstrate the accuracy of KSSOLV. These systems are relatively small and can be solved between tens of seconds to tens of minutes. The number of iterations required to reach convergence and the wall clock time it takes generally increase with the system size. The largest system we tested is the Si$_{216}$ cluster with $216$ atoms and $864$ electrons. More than 1 million plane-waves are used in this calculation that employs the PBE functional. The entire calculation took more than two hours.

The HSE06 functional is used for Si$_{64}$ and C$_{60}$ to perform hybrid functional calculations for these systems with and without using the ACE method. We can see from Table~\ref{tab:benchset} that,  without using ACE, the hybrid functional calculations for these systems are two orders of magnitude more expensive than the corresponding PBE calculations for the same systems. When ACE is used, the hybrid functional calculations are only $4\sim 5$ times more expensive than the corresponding PBE calculations. In previous studies ~\cite{kssolvgpu}, we also compared different diagonalization algorithms.

In addition to insulators and semiconductors, we also measure the performance of KSSOLV on a metallic system Cu$_4$. The calculation, which uses 64 k-points, can be completed in less than 10 minutes.

\begin{figure*}[ht]
    \centering
    \includegraphics[width=1\textwidth]{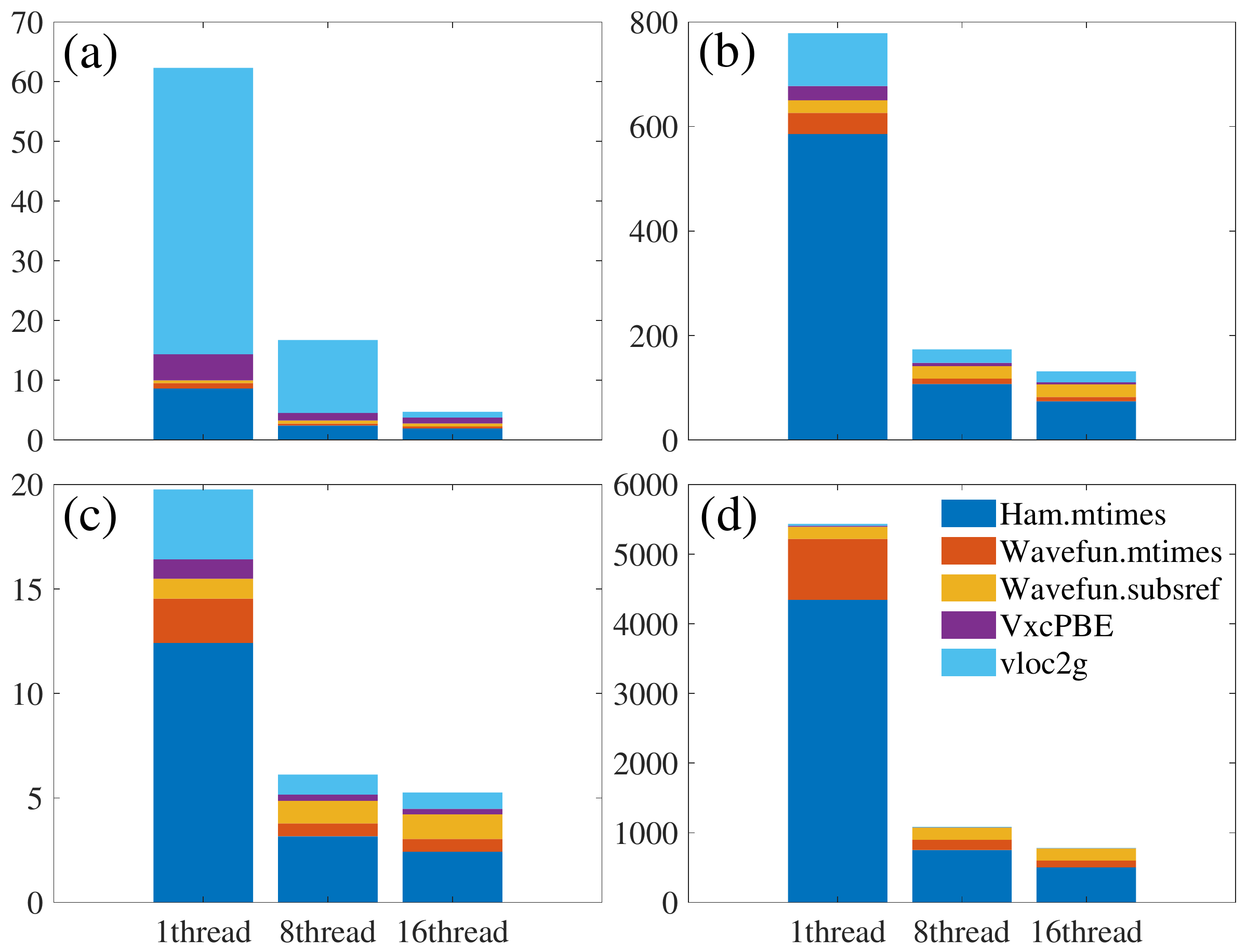}
    \caption{The profiling results of different systems PBE-calculation within 1, 8, 16 threads, including Ham.mtimes, Wavefun.mtimes, Wavefun.subsref, VxcPBE, vlov2g. Systems: (a) silane (SiH$_{4}$) molecule, (b) benzene (C$_{6}$H$_{6}$) molecule, (c) bulk silicon Si$_{8}$, (d) bulk silicon Si$_{64}$. All of these results come from the whole calculation process.}
    \label{fig:multical}
\end{figure*}

Because MATLAB can take advantage of  multiple threads on a many-core CPU to parallelize many computational kernels, we can speed up KSSOLV calculation on such a CPU without additional parallelization effort. In Figure~\ref{fig:multical}, we report the parallel scaling of KSSOLV when it is used to compute the ground states of four different systems using the PBE functional. We observe that the total wall clock time can be reduced by a factor of 5 when the number of threads is increased from 1 to 8. Increasing the number of threads further to 16 leads to an additional reduction in wall clock time. However, the reduction factor is much smaller. By default, MATLAB generally will try to use the maximum number of threads available on the machine being used. As a result, KSSOLV can benefit from the maximum shared memory concurrency available on any many-core CPUs.

In addition to reporting the total wall clock time, we also show a breakdown of timing among several key computational components of KSSOLV. The Hamiltonian wavefunction multiplication (labelled by {\tt Ham.mtimes}), which performs {\tt HX = H*X} as an overloaded matrix-matrix multiplication between a {\tt Ham} object {\tt H} and a {\tt Wavefun} object {\tt X} as explained in section~\ref{sec:hamclass}, constitutes the largest cost. This is followed by the cost of {\tt  Wavefun.mtimes} which performs dense matrix-matrix multiplications between two {\tt Wavefun} objects or between a {\tt Wavefun} object and a regular matrix.  The {\tt Wavefun.subsref} function, which is used to extract a subset of wavefunctions, involves mainly data movement and copying. Such data movement cannot be easily parallelized. Hence, the timing associated with {\tt Wavefun.subsref} does not decrease as the number of threads increases. The {\tt VxcPBE} function, which is used to evaluate the PBE exchange-correlation energy and potential, takes a small fraction of the time. The function {\tt vlov2g} is used to convert local pseudopotential on a non-uniform grid in real-space to a uniform grid in the reciprocal space in the initialization of the pseudopotential. This one-time cost can be relatively large for small systems, but becomes negligible when the system size becomes sufficiently large. 

\section{Conclusion and outlook} \label{sec:Conclusion}

KSSOLV 2.0 preserves the main object-oriented design features of the original KSSOLV software toolbox for solving the Kohn-Sham equations. Such design features make it easy for users to set up a problem and obtain a solution. They also enable developers to easily prototype and test new algorithms.  The new version contains more advanced algorithms such as ACE, PC-DIIS, ISDF for hybrid functional DFT calculations, and new functionalities such as geometry optimization.  The software produces accurate results that are consistent with those produced by other plane-wave based KS-DFT software such as QE.  It is efficient for performing KS-DFT electronic structure calculations for small to medium sized problems. It is a great teaching tool that can help students and researchers quickly learn how to analyze the electronic structure of molecules and solids. At the same time, it can also be a useful research tool in chemical and materials sciences for analyzing properties of interesting materials or chemical systems and for developing more efficient numerical methods.
 Although KSSOLV 2.0 is designed to perform ground state DFT  and geometry optimization calculations, several new developments are already underway to include new functionalities in the next release. In particular, we plan to include functionalities to allow users to perform time-dependent density functional theory (TDDFT) calculations and post-DFT calculations such as computing GW\cite{gw} quasi-particle energies and eigenvalues and eigenvectors of the Bethe-Salpeter Hamiltonian~\cite{bse}. In addition, we will integrate KSSOLV with machine learning tools to accelerate the materials design and discovery process.





\vspace{3ex}

\section*{Acknowledgments}

This work is partly supported by the National Natural Science Foundation of China under the Grant No. 22173093 (W.H.), 22073086 (L.J.), 21803064 (Z.D.) and 21688102 (J.Y.), by the School of Future Technology under Grant No. KF2020003 (W.H.), by the Chinese Academy of Sciences Pioneer Hundred Talents Program under Grant No. KJ234007002 (W.H.), by the National Key Research and Development Program of China under the Grant No. 2016YFA0200604 (J.Y.), the Anhui Initiative in Quantum Information Technologies under Grant No. AHY090400 (J.Y.), the Center of Chinese Academy Project for Young Scientists in Basic Research under Grant No. YSBR-005 (W.H.), the School of Future Technology under Grant No. SK2340002001 (W.H.), the Fundamental Research Funds for the Central Universities under Grant No. WK2340000091 (W.H.) and WK2060000018 (W.H.) from University of Science and Technology of China. This work was also supported by the Scientific Discovery through Advanced  Computing (SciDAC) program and the Center for Applied Mathematics for Energy Research Applications (CAMERA) funded by U.S. Department of Energy, Office of Science, Advanced Scientific Computing Research under Contract No. DE-AC02-05CH11231 (L.L., C.Y.). The authors thank the Hefei Advanced Computing Center, the Supercomputing Center of Chinese Academy of Sciences, the Supercomputing Center of USTC and the National Energy Scientific Computing Center (NERSC) for the computational resources. We would like to thank Temo Vekua, Reza Fazel-Rezai, Kajia Ruan and Wei Chen at MathWorks for helpful suggestions.

\bibliographystyle{elsarticlenum.bst}
\bibliography{reference}

\end{document}